\def\version{March 2, 2001}

\documentclass[reqno,11pt]{amsart}

\usepackage{epsf}

\def\be{\begin{equation}}
\def\ba{\begin{align}}
\def\bm{\begin{multline}}
\def\bfig{\begin{figure}[htb]}
\def\efig{\end{figure}}

\newcommand{\bibit}[1]{\bibitem[#1]{#1}}
\newcommand{\paper}[1]{{\it #1}, }
\newcommand{\journal}[4]{#1 {\bf #2}, #3 (#4)}
\newcommand{\CMP}{Commun. Math. Phys.}
\newcommand{\HPA}{Helv. Phys. Acta}
\newcommand{\JSP}{J. Stat. Phys.}
\newcommand{\PR}{Phys. Rev.}

\newcommand{\PRB}{Phys. Rev. B}
\newcommand{\PRL}{Phys. Rev. Lett.}

\newcommand{\JPA}{J. Phys. A}
\newcommand{\JMP}{J. Math. Phys.}
\newcommand{\LMP}{Lett. Math. Phys.}

\numberwithin{equation}{section}
\newtheorem{theorem}{Theorem}[section]

\newcommand{\fig}{Fig.\;}

\newcommand{\eg}{e.g.\;}
\newcommand{\ie}{i.e.\;}

\newcommand{\nn}{\nonumber}
\def\bbbone{{\mathchoice {\rm 1\mskip-4mu l} {\rm 1\mskip-4mu l} {\rm
1\mskip-4.5mu l} {\rm 1\mskip-5mu l}}}

\DeclareMathSymbol{\leqslant}{\mathalpha}{AMSa}{"36}
\DeclareMathSymbol{\geqslant}{\mathalpha}{AMSa}{"3E}
\DeclareMathSymbol{\doteqdot}{\mathalpha}{AMSa}{"2B}
\DeclareMathSymbol{\circlearrowright}{\mathalpha}{AMSa}{"08}
\DeclareMathSymbol{\subsetneq}{\mathalpha}{AMSb}{"28}
\DeclareMathSymbol{\supsetneq}{\mathalpha}{AMSb}{"29}
\renewcommand{\leq}{\;\leqslant\;}
\renewcommand{\geq}{\;\geqslant\;}

\newcommand{\dd}{{\rm d}}
\newcommand{\e}[1]{\,{\rm e}^{#1}\,}
\newcommand{\ii}{{\rm i}}
\newcommand{\sumtwo}[2]{\sum_{\substack{#1 \\ #2}}}

\newcommand{\uniontwo}[2]{\union_{\substack{#1 \\ #2}}}
\newcommand{\inttwo}[2]{\int_{\substack{#1 \\ #2}}}

\newcommand{\limtwo}[2]{\lim_{\substack{#1 \\ #2}}}
\DeclareMathOperator*{\union}{\text{\large$\cup$}}
\DeclareMathOperator*{\inter}{\text{\large$\cap$}}

\def\Tr{{\operatorname{Tr\,}}}

\def\dist{{\operatorname{dist\,}}}

\def\bra #1{\langle#1 |\,}
\def\ket #1{\,|#1 \rangle}

\newcommand{\expval}[1]{\langle #1 \rangle}

\newcommand{\compl}{{\rm c}}
\newcommand{\indicator}[1]{\,\mathbb I \, \bigl[ #1 \bigr]}

\def\writefig#1 #2 #3 {\rlap{\kern #1 truecm \raise #2 truecm
\hbox{#3}}}
\def\figtext#1{\smash{\hbox{#1}} \vspace{-5mm}}

\newcommand{\caA}{{\mathcal A}}
\newcommand{\caB}{{\mathcal B}}

\newcommand{\caF}{{\mathcal F}}

\newcommand{\caH}{{\mathcal H}}

\newcommand{\caN}{{\mathcal N}}

\newcommand{\bbB}{{\mathbb B}}
\newcommand{\bbC}{{\mathbb C}}

\newcommand{\bbE}{{\mathbb E}}

\newcommand{\bbN}{{\mathbb N}}

\newcommand{\bbR}{{\mathbb R}}

\newcommand{\bbZ}{{\mathbb Z}}

\newcommand{\bsn}{{\boldsymbol n}}

\newcommand{\bsx}{{\boldsymbol x}}

\newcommand{\bsA}{{\boldsymbol A}}
\newcommand{\bsB}{{\boldsymbol B}}

\newcommand{\bsLambda}{{\boldsymbol \Lambda}}

\begin{document}

\begin{quote}
\raggedleft
{\small
\version
}
\end{quote}
\vspace{2mm}

\title[Geometric \& probabilistic aspects of boson lattice models]{Geometric and
probabilistic aspects \\ of boson lattice models}

\author{Daniel Ueltschi}

\address{Daniel Ueltschi \hfill\newline
Department of Physics\hfill\newline
Princeton University\hfill\newline
Jadwin Hall\hfill\newline
Princeton, NJ 08544\hfill\newline
{\small\rm\indent http://www.princeton.edu/$\sim$ueltschi}}
\email{ueltschi@princeton.edu}

\maketitle

\vspace{-5mm}

\begin{centering}
{\small\it
Department of Physics, Princeton University, New Jersey\\
}
\end{centering}

\vspace{5mm}

\begin{quote}
{\small
{\bf Abstract.}
This review describes quantum systems of bosonic particles moving on a lattice. These
models are relevant in statistical physics, and have natural ties with probability theory.
The general setting is recalled and the main questions about phase transitions are
addressed. A lattice model with Lennard-Jones potential is studied as an example of a
system where first-order phase transitions occur.

A major interest of bosonic systems is the possibility of displaying a Bose-Einstein
condensation. This is discussed in the light of the main existing rigorous result, namely
its occurrence in the hard-core boson model. Finally, we consider another approach that
involves the lengths of the cycles formed by the particles in the space-time
representation; Bose-Einstein condensation should be related to positive probability of infinite cycles.
\vspace{1mm}

}  

\end{quote}

\renewcommand{\thefootnote}{}
\footnote{Work partially supported by the US National Science Foundation, grant
PHY-98 20650.}
\setcounter{footnote}{0}
\renewcommand{\thefootnote}{\arabic{footnote}}

\vspace{3mm}

\section{Introduction}

Statistical Physics is the study of macroscopic properties of systems with a large number
of microsopic particles. Its relevance stems from the law of large numbers, allowing the
state of a system to be specified by the values of a few `macroscopic variables', although
the number of microscopic degrees of freedom is enormous. From a probability theory point of view, the Ising
model of classical spins is an example of identically distributed, but not independent,
random variables; when couplings are small (high temperature, random variables
close to independent), magnetization is zero; for
large couplings however (strong dependence, or low temperature), the law of large numbers
takes a subtler form, with two typical values for the magnetization. This behavior is a
manifestation of a phase transition. Connections between statistical physics and
probability theory, such as the relation between the physical entropy and the rate function
of large deviations, are discussed in detail by Pfister in his excellent lectures \cite{Pfi}.

While the original motivation for the Ising model resides in quantum mechanics, it is
considered as a classical model, because energy and observables are functions on the space
of configurations --- in quantum systems, these are operators on the vector space spanned
by the configurations.
There are several reasons for devoting some attention to quantum systems.
\begin{itemize}
\item They are closer to the physical reality,
and usually of more interest to physicists than classical ones.
\item They have richer properties; new types of phases such as superfluidity or superconductivity
may show up that are intrisically quantum phenomena.
\item They pose a number of mathematically interesting questions.
\end{itemize}

There are three classes of quantum lattice systems. The first class consists of spin systems, such as the quantum
Heisenberg model, where each site of the lattice hosts a spin that interacts with nearest
neighbors. In the second class are fermionic systems, an example of which is the Hubbard model, where the
energy of the quantum particles is provided by a discrete Laplacian (`hopping matrix') for
the kinetic part, while the potential part is given by an operator that is a function of
the position operators; particles are indistinguishable, so that a permutation of the
particles results in the same quantum state, up to a sign for odd permutations. The last
class consists in
bosonic systems that describe particles hopping on a lattice and interacting among
themselves, but a permutation does not alter their wavefunction. There are also other
models that have spins and particles, particles with spins, or both kinds of particles.

This review is focussed on bosonic systems. Their great advantage over fermionic ones is that
they involve only positive numbers, hence natural links with probability theory. They also
have extremely interesting behavior with various phase transitions, including the
Bose-Einstein condensation (hereafter denoted BEC), that should be one of the mechanisms
leading to
superfluidity and superconductivity.

Section \ref{secmathst} introduces the general formalism and defines equilibrium states.
This leads to the notion of phase transitions, and of symmetry breaking. These ideas are
then illustrated in a simple boson model with Lennard-Jones potential; its low temperature
phase diagram is analyzed and shown to display various phase transitions (Section
\ref{secexample}). This can be proven by showing the equivalence of this model with a
`contour model' that fits the framework of the Pirogov-Sinai theory (Section
\ref{seccontrep}). These techniques, however useful, do not allow discussing the occurrence
of BEC. We briefly review the main questions in Section \ref{secBEC}, and state the best
result so far --- the occurrence of `off-diagonal long-range order' in the hard-core boson
lattice model \cite{DLS,KLS}, see Theorem \ref{thmBEChcb}. We conclude by discussing an
approach to the BEC that is both geometric and probabilistic, and that involves `cycles'
formed by bosonic trajectories in the Feynman-Kac representation. When the temperature
decreases, the probability of observing an infinite cycle should vary from 0 to a positive
number, and this transition should be related to BEC. These ideas are described in Section
\ref{secinfcyc}.

\section{Mathematical structure}
\label{secmathst}

\subsection{Microscopic description}

The physical picture is that of a group of bosons on a lattice, with the kinetic energy described by a
discrete Laplacian, and interacting with a two-body potential.

Let $\Lambda \subset \bbZ^d$ be a finite volume. The space $\bbC^\Lambda$ of `wave
functions' on $\Lambda$ is a Hilbert space, and a
normalized vector describes the state of a quantum particle. For $\Psi \in \otimes_{n=1}^N
\bbC^\Lambda$ we define the symmetrization operator $S_N$
$$
S_N \Psi(x_1, \dots, x_N) = \frac1{N!} \sum_\pi \Psi(x_{\pi(x)}, \dots, x_{\pi(N)}),
$$
where the sum is over all permutations of $N$ elements.
Then $S_N(\otimes_{n=1}^N \bbC^\Lambda)$ is the Hilbert space for $N$ bosonic particles,
and the Fock space that describes a variable number of particles is $\caF_\Lambda =
\oplus_{N=0}^\infty S_N(\otimes^N \bbC^\Lambda)$. There is a natural inner product on this
space that makes it into a Hilbert space.

This formalism is the natural one from a physical point of view, but it is more practical
to consider another Hilbert space that is isomorphic to the Fock space above. Thus we start
again, this time in the appropriate setting. Standard references are Israel \cite{Isr} and
Simon \cite{Sim}.

We consider a Hilbert space $\caH_0$; either $\caH_0 \simeq \bbC^\infty$ (more precisely
$\caH_0 \simeq \ell^2(\bbC)$), or $\caH_0 \simeq
\bbC^N$ for systems with a `hard-core condition', \ie a prescription that sets a maximal
number $N$ of bosons at a given site. Then we define local Hilbert spaces $\{\caH_x\}_{x\in\bbZ^d}$ with each
$\caH_x \simeq \caH_0$, and for
$\Lambda \subset \bbZ^d$ we set $\caH_\Lambda = \otimes_{x \in \Lambda} \caH_x$.

A natural basis for $\caH_0$ is $\{ \ket{n_0} \}_{n_0 \in \bbN}$; for $\caH_\Lambda$, an
element of this basis is
\be
\label{defbasis}
\ket n = \otimes_{x \in \Lambda} \ket{n_x},
\end{equation}
where $n \in \bbN^\Lambda$. This represents a state where the site $x$ has $n_x$ bosons. The main
operators are the {\it creation operator} of a boson at site $x$, noted $c_x^\dagger$, its
adjoint the {\it annihilation operator} $c_x$, and the {\it operator number of particles}
at $x$, $\hat n_x = c_x^\dagger c_x$. Their actions on the above basis are
\ba
c_x^\dagger \ket n &= \sqrt{n_x+1} \ket{n+\delta_x}, \nn\\
c_x \ket n &= \sqrt{n_x} \ket{n-\delta_x}, \\
\hat n_x \ket n &= n_x \ket n. \nn
\end{align}
Here, we denoted $\ket{n+\delta_x}$ the vector that is equal to $\otimes_{y\in\Lambda}
\ket{n_y+\delta_{xy}}$. Considering a system with hard-core bosons, we demand that $c_x^\dagger \ket n
= 0$ if $n_x = N$. Notice that the operators $\hat n_x$ are diagonal in this basis.
Without hard-cores, creation and annihilation operators satisfy the commutation relations
\be
[c_x, c_y^\dagger] = \delta_{xy}.
\end{equation}
With a hard-core, the relation is
\be
[c_x,c_y^\dagger] = \delta_{xy} \Bigl\{ 1 - (N+1) \sum_{n: n_x=N} \ket n \bra n \Bigr\}.
\end{equation}
In order to avoid extra technicalities associated with unbounded operators, we restrict our
interest to models with a hard-core condition.

The energy of the particles is given by an `interaction', that is, a collection of
operators $H = (H_A)_{A \subset \bbZ^d}$ with $H_A : \caH_A \to \caH_A$. We commit an abuse
of notation and still denote $H_A$ the operator $H_A \otimes \bbbone_{\Lambda\setminus A}$.
We define
operations $(H+H')_A = H_A + H_A'$ and $(\lambda H)_A = \lambda H_A$, and introduce the norm
\be
\label{defnorm}
\|H\|_r = \sup_x \sum_{A \ni x} \|H_A\| \e{r \|A\|}
\end{equation}
for some positive number $r$, where $\|A\|$ is the cardinality of the smallest connected set
that contains $A$. An interaction is periodic iff
there exists a subgroup $\Lambda' \subset \bbZ^d$ of dimension $d$, such that $H_{\tau_x A}
= H_A$ for all $x \in \Lambda'$. Here, $\tau_x$ is the translation operator. The space of {\it periodic} interactions with finite norm
\eqref{defnorm} is a Banach space and we denote it $\caB$.

\subsection{Free energy and equilibrium states}

The {\it free energy}\footnote{Some authors prefer to define the {\it
pressure} instead, that is equal to $-\beta$ times the free energy. In thermodynamics, the
pressure is the potential depending on temperature, volume, and chemical potential. It
would be physically more appropriate for the discussion of boson models below. The free
energy is however more convenient for low temperature studies, since $\lim_{\beta\to\infty}
f(H)$ exists in typical situations.} for an interaction $H$ and at inverse
temperature $\beta$ is
\be
\label{deffen}
f(H) = -\frac1\beta \lim_{\Lambda \nearrow \bbZ^d} \frac1{|\Lambda|} \log \Tr \e{-\beta
\sum_{A \subset \Lambda} H_A},
\end{equation}
where the limit is taken over a sequence $(\Lambda_n)$ of volumes such that $\lim_n
\frac{|\partial_r \Lambda|}{|\Lambda|} = 0$ for all $r$; here, $\partial_r \Lambda = \{x \in \Lambda :
\dist(x,\Lambda^\compl) \leq r\}$ is an enlarged boundary of $\Lambda$. It is well-known that the
limit \eqref{deffen} is independent of the way the limit is performed, and that it is a
concave function of the interactions.

An {\it equilibrium state} $\rho_H$ for the interaction $H$ is a linear, normalized,
positive functional on the space of interactions, that is tangent to the free energy at
$H$, \ie for all $K \in \caB$,
\be
\label{defstate}
\rho_H(K) + f(H) \geq f(H+K).
\end{equation}

To motivate this definition, let us consider the free energy at finite volume
$f_\Lambda(H)$, given by \eqref{deffen} without taking the limit. The corresponding finite-volume
state would be
$$
\rho^\Lambda_H(K) = \frac\dd{\dd\lambda} f_\Lambda(H+\lambda K) \Big|_{\lambda=0} =
\frac{\Tr(\frac1{|\Lambda|} \sum_{A \subset \Lambda} K_A) \e{-\beta \sum_{A \subset
\Lambda} H_A}}{\Tr \e{-\beta \sum_{A \subset \Lambda} H_A}}.
$$

The definition \eqref{defstate} is therefore more general, and allows to define states
directly with the free energy in the limit of infinite volumes. The set of tangent
functionals at a given $H$ is convex; extremal points are the `pure states'. Existence of more than one
tangent functionals implies a {\it first-order phase transition}.

A popular definition of equilibrium states in quantum lattice systems involves `KMS
states'. They are actually equivalent to tangent functionals, see \eg \cite{Isr,Sim}.

One could restrict our interest to operators that are diagonal with respect to the basis
\eqref{defbasis}
above. In this case, one would consider the configuration space $\bbN^\Lambda$ and the
interactions would be collections of functions on this space. As a result, we have a
classical system, whose free energy is still given by \eqref{deffen}. States can also be
defined as tangent functionals to the free energy.

Hamiltonians (or interactions, in our case) may possess {\it symmetries}: for instance, a
translation by a vector of the lattice often does not affect the energy, nor does a rotation or
a reflection. In quantum statistical physics, one says that $U: \caB \to \caB$, $H \mapsto
H' = U(H)$ is a symmetry if for all volumes $\Lambda$ that appear in the limit in \eqref{deffen} there
exists a unitary operator $U_\Lambda$ in $\caH_\Lambda$ such that
\be
\label{defsym}
U_\Lambda \sum_{A \subset \Lambda} H_A \, U_\Lambda^{-1} = \sum_{A\subset\Lambda} H_A'.
\end{equation}
Clearly, one has $f(H') = f(H)$.

Let us illustrate this notion on two examples that will be relevant in the sequel. The
first one is the translation by one site in the direction 1; it is defined by $H_A' =
H_{A-e_1}$, where $A-e_1 = \{x: (x(1)+1, x(2), \dots, x(d)) \in A\}$. Let us assume that the
boxes $\Lambda$ are rectangles with periodic boundary conditions, and $1 \leq x(1) \leq L$.
Then one can choose $U_\Lambda$ to be
$U_\Lambda \ket{n_\Lambda} = \ket{n_\Lambda'}$, where $n_x' = n_{(x(1)-1, x(2), \dots, x(d))}$
if $x(1) \neq 1$, $n_x' = n_{(L, x(2), \dots, x(d))}$ if $x(1)=1$.

The second example is relevant for the Bose-Einstein condensation and is called a `global gauge
symmetry'; $U_\Lambda$ takes the form $U_\Lambda = \e{\ii\alpha \sum_{x\in\Lambda} \hat
n_x}$, $\alpha \in [0,2\pi)$. Hamiltonians describing real particles always conserve the total number of
particles, and hence possess the global gauge symmetry. It can be broken however, yielding states
where the number of particles fluctuates more than usual.\footnote{Large deviations of the
number of particles in a finite volume are studied in \cite{LLS} in the ideal Bose gas,
outside the condensation regime. They are indirectly affected by BEC, if the deviated phase
is a condensate.} We discuss this in Section \ref{secBEC}.

\section{Example: Hopping particles with two-body interactions}
\label{secexample}

In this section we introduce a simple lattice model and study it by means of geometric
methods. One obtains that the free energy display angles corresponding to first-order phase
transitions, see \fig\ref{figgse} below. Let us mention that the existence of a first-order
phase transition in a quantum
system {\it in the continuum} has been recently established for the (quantum) Widom-Rowlinson model \cite{CP,Iof}.

\subsection{The model}

The particles have kinetic and potential energy, so that the Hamiltonian is
\be
\label{defHam}
H = T + V.
\end{equation}
The kinetic energy $T$ of particles on a lattice is described by a discrete Laplacian that can
be written using the creation and annihilation operators in the following way: $T = (T_A)$,
with
\be
T_A = \begin{cases} -t (c_x^\dagger c_y + c_y^\dagger c_x) & \text{if } A = \{x,y\} \text{ with
} |x-y|=1 \\ 0 & \text{otherwise.} \end{cases}
\end{equation}

We consider here two-body interactions given by a function $U(\cdot)$ that depends on the
Euclidean
distance between two particles.
\be
V_A = \begin{cases} U(|x-y|) \, \hat n_x \hat n_y & \text{if } A = \{x,y\} \text{ and } x \neq
y \\ \tfrac12 U(0) \, \hat n_x (\hat n_x-1)  & \text{if } A = \{x\} \\ 0 &
\text{otherwise.} \end{cases}
\end{equation}
The on-site operator $\frac12 \hat n_x (\hat n_x-1)$ is the number of pairs of particles at
site $x$, and the energy is naturally proportional to it. The model with only on-site
interactions was introduced in \cite{FWGF} and is usually called the Bose-Hubbard
model.

In order for the Hamiltonian $H=T+V$ to have finite norm \eqref{defnorm}, the interaction $U$ must have
exponential decay for large distances. The density of the system is controlled by a term
involving a chemical potential, $-\mu N$, where $N$ is the `interaction' that corresponds
to the number of particles; $N_{\{x\}} = \hat n_x$ and $N_A = 0$ if $|A|\geq2$.

Let us now discuss in more details the case of a Lennard-Jones type of potential; the graph
of the corresponding $U$ is depicted in \fig\ref{figpot}. We suppose that $U(0)=+\infty$,
corresponding to a hard-core condition that prevents multiple occupancy of the sites. We
also suppose that the tail
\be
u_r = \sum_{|y| \geq 2} |U(|y|)| \e{r|y|}
\end{equation}
does not play an important role; only important values of the potential are $U(1)$ and
$U(\sqrt2)$. The results below are valid for $u_r \leq u_0$, the values of $u_0$ and $r$ depending on
$U(1)$ and $U(\sqrt2)$.

\bfig
\epsfxsize=70mm
\centerline{\epsffile{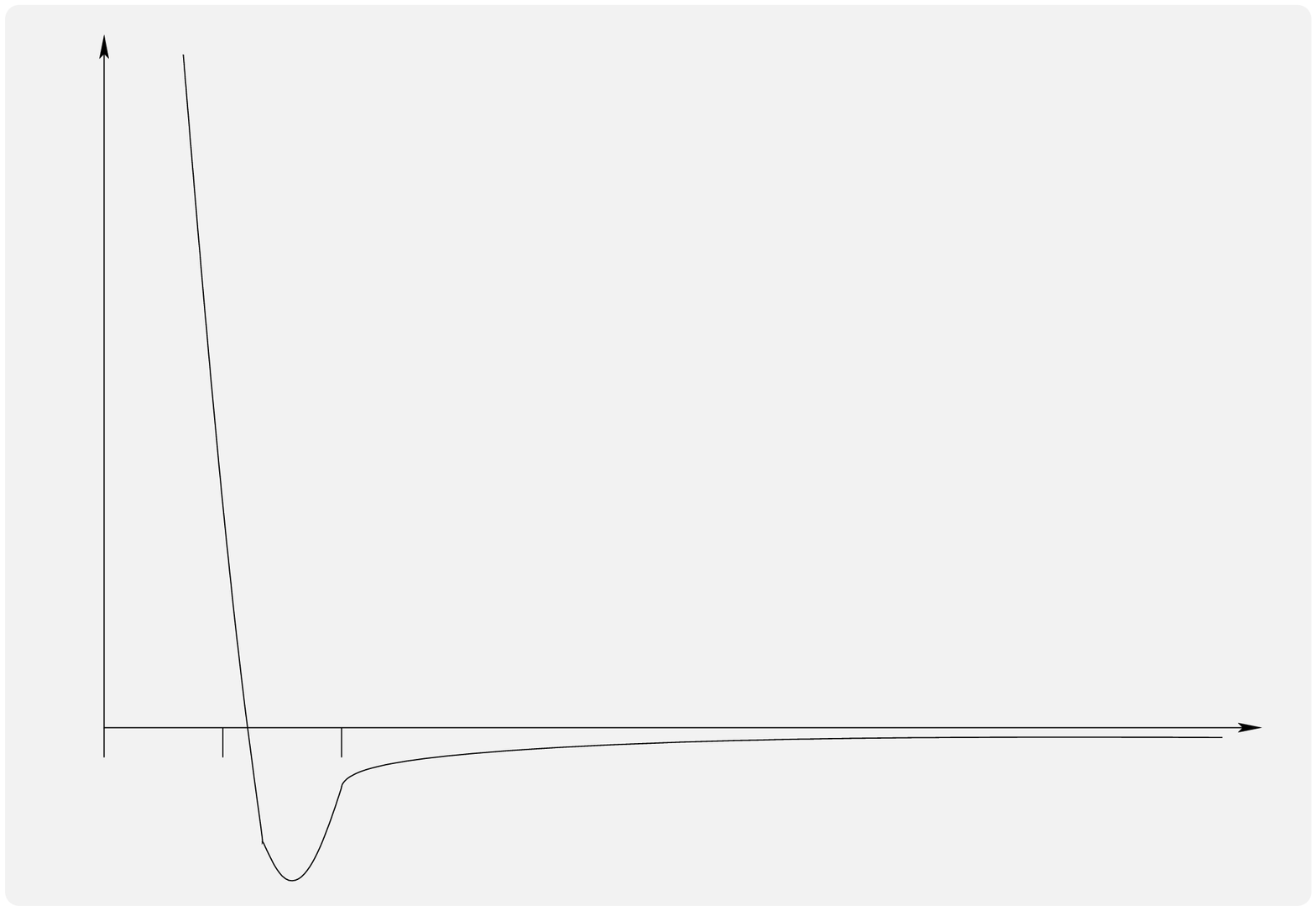}}
\figtext{
\writefig   3.2  1.3  {\small $a$}
\writefig   -3.2  5.3  {\small $U(a)$}
\writefig   -2.95  1.1  {\tiny $0$}
\writefig  -2.33 1.1  {\tiny $1$}
\writefig   -1.75  1.1  {\tiny $2$}
}
\caption{The graph of a Lennard-Jones type of potential.}
\label{figpot}
\end{figure}

We start with an analysis of the ground states of the `classical model' with configuration
space $\{0,1\}^\Lambda$ and a Hamiltonian given as a sum over squares $S$ of four
nearest-neighbor sites:
\be
H_\Lambda^{\text{cl}}(n) = \sum_{S \subset \Lambda} \Bigl[ \frac{U(1)}{2(d-1)} \sumtwo{\{x,y\}
\subset S}{|x-y|=1} n_x n_y + U(\sqrt2) \sumtwo{\{x,y\} \subset S}{|x-y|=\sqrt2} n_x n_y
\Bigl] - \frac14 \sum_{x \in S} \bigl[ \mu \, n_x + h \, (-1)^x n_x \bigr].
\end{equation}
We added a staggered interaction $-h (-1)^x n_x$, with $(-1)^x \equiv
(-1)^{\|x\|_1}$. This interaction has no physical relevance, but is mathematically
useful to uncover the occurrence of phases of the chessboard type that breaks the symmetry
of translation invariance. One is of course interested in what happens when $h=0$.

Four configurations are important, namely $(\begin{smallmatrix}0&0\\0&0\end{smallmatrix})$,
$(\begin{smallmatrix}1&0\\0&1\end{smallmatrix})$,
$(\begin{smallmatrix}0&1\\1&0\end{smallmatrix})$,and
$(\begin{smallmatrix}1&1\\1&1\end{smallmatrix})$; respective energies are
\ba
\label{gsen}
e^{\mu,h}(\begin{smallmatrix}0&0\\0&0\end{smallmatrix}) &= 0 \nn\\
e^{\mu,h}(\begin{smallmatrix}1&0\\0&1\end{smallmatrix}) &= U(\sqrt2) - \tfrac\mu2 - \tfrac
h2 \nn\\
e^{\mu,h}(\begin{smallmatrix}0&1\\1&0\end{smallmatrix}) &= U(\sqrt2) - \tfrac\mu2 + \tfrac
h2 \\
e^{\mu,h}(\begin{smallmatrix}1&1\\1&1\end{smallmatrix}) &= 2U(1) + 2U(\sqrt2) - \mu. \nn
\end{align}

We make the further assumptions on the potential that $U(1)>0$,
ensuring a chessboard phase to be present, and $U(\sqrt2)<0$, so that no phases with quarter
density show up --- they are more difficult to study, since the classical model has an
infinite number of ground states. In many cases one expects that this degeneracy will be
lifted as a result of `quantum fluctuations', that is, the effect of a small kinetic energy
$T$. A general theory of such effects combined with the Pirogov-Sinai theory can be found
in \cite{DFFR,KU}. Notice that $U(1)>U(\sqrt2)$, meaning that at low temperature, the
chessboard phase overcomes the phase with alternate rows or columns of 1's and
0's. Energies \eqref{gsen} provide
the zero-temperature phase diagram and allow guesses for the low temperature situation.

\subsection{The phase diagram}

The situation at high temperature ($\beta$ small) is that of bosons with weak interactions
and no phase transitions may occur. The natural condition for high temperature is that
$\beta \|H\|_r$ is small; one can however prove slightly more by {\it not} requesting
that $U(0)$ be small. So we define (compare with \eqref{defnorm})
\be
\|H\|_r^* = \sup_x \sumtwo{A \ni x}{|A| \geq 2} \|H_A\| \e{r \|A\|}.
\end{equation}

\begin{theorem}
\label{thmht}
There exists $r<\infty$ such that if $\beta \|H\|_r^* < 1$, there is a
unique tangent functional at $H$, and the free energy is real analytic in a neighborhood of
$H$.
\end{theorem}

This theorem is proven in Section \ref{subsecproofht} using high temperature expansions. We shall see below that there
may be more than one tangent functionals at low temperature, corresponding to equilibrium states
that are not translation invariant. This implies that a transition with symmetry breaking
takes place when the temperature decreases. Presumably it is second order, like in the
Ising model, but there are no rigorous results to support this.

The limit $\beta\to\infty$ is easily analyzed and is depicted in \fig\ref{figgse}.
\bfig
\epsfxsize=80mm
\centerline{\epsffile{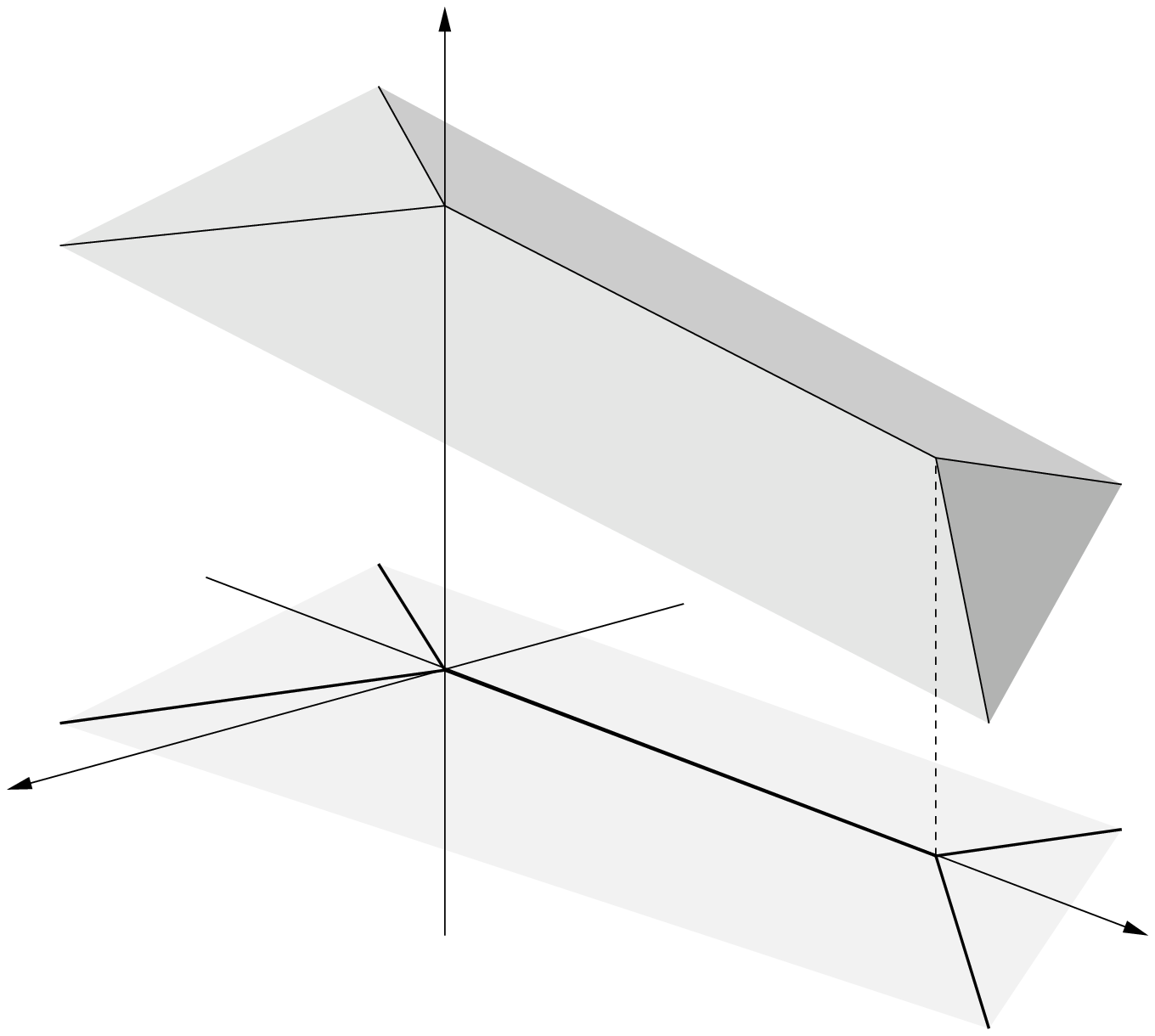}}
\figtext{
\writefig   3.8  1.1  {\small $\mu$}
\writefig   -4.1  2.0  {\small $h$}
\writefig   -0.7  7.4  {\small $e^{\mu,h}$}
\writefig   -2.6  3.1  {\footnotesize $(\begin{smallmatrix}0&0\\0&0\end{smallmatrix})$}
\writefig   1.1  2.5  {\footnotesize $(\begin{smallmatrix}0&1\\1&0\end{smallmatrix})$}
\writefig   -0.5  2.1  {\footnotesize $(\begin{smallmatrix}1&0\\0&1\end{smallmatrix})$}
\writefig   2.7  1.1  {\footnotesize $(\begin{smallmatrix}1&1\\1&1\end{smallmatrix})$}
}
\caption{The free energy in the limit $\beta\to\infty$. The phase diagram is divided in
four domains, corresponding to the empty, chessboards, and full configurations. For large
$\beta$ and small $t$, the flat parts bend but the angles remain.}
\label{figgse}
\end{figure}
The graph of the function $e^{\mu,h}$ is a kind of roof with four flat parts. There
are angles between each flat part, so that first derivatives have discontinuities there. The
two questions that should be asked are:
\begin{itemize}
\item Does this picture survive when adding the tail of the potential, and the kinetic
energy (hopping matrix)?
\item Does this picture survive at non-zero temperatures?
\end{itemize}

The answer to both questions is yes and is provided by the {\it quantum Pirogov-Sinai
theory}. It can be viewed as a considerable extension of the Peierls argument for the Ising
model. It was proposed by Pirogov and Sinai for classical lattice models \cite{PS,Sin}, and
extended to quantum models in \cite{BKU,DFF,DFFR,KU,FRU}. These ideas are discussed for
this model in the next section. One is then led to the phase diagram of \fig\ref{figphd}.

Multiple phases and occurrences of first order phase transitions are proven when $\beta$ is
large and $t$ small, \ie at low temperature and close to the classical limit of
vanishing hoppings. It is expected that BEC and superfluidity are present in dimension $d
\geq 3$, when the temperature is low and with sufficient hoppings \cite{FWGF}. Actually,
the situation $U(0)=\infty$ and $U(a)=0$ for $a \geq 1$ corresponds to the hard-core boson
model, when BEC is proven at low temperature \cite{DLS,KLS}; see Section \ref{secBEC}.

\bfig
\epsfxsize=80mm
\centerline{\epsffile{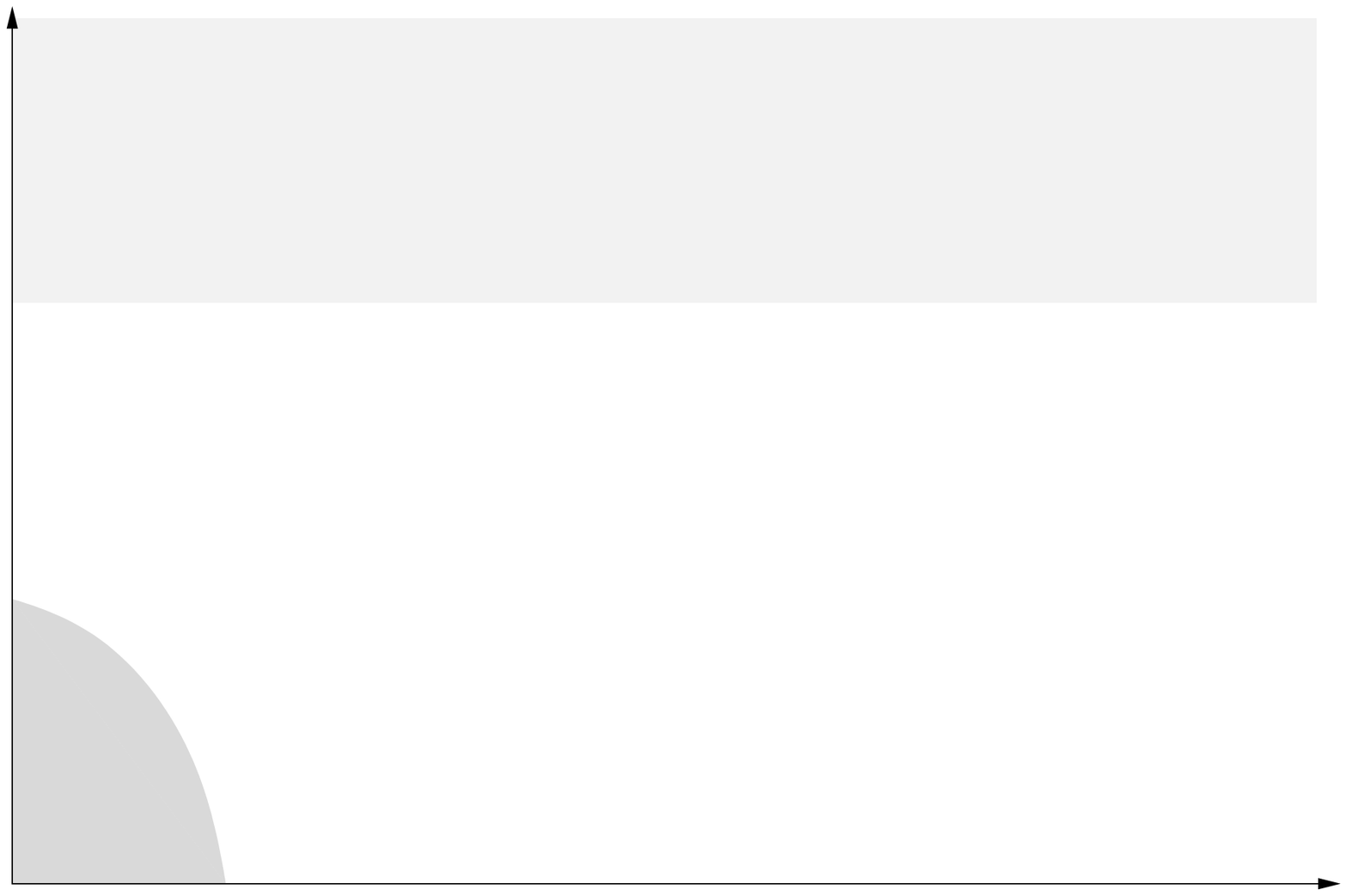}}
\figtext{
\writefig   3.9  0.5  {\small $t$}
\writefig   -4.3  5.5  {$\frac1\beta$}
\writefig   -1  4.7  {\footnotesize Unicity}
\writefig   -3.7  1.1  {\footnotesize LRO}
\writefig   -0.5  1.5  {\footnotesize BEC expected}
}
\caption{The phase diagram $(t,\frac1\beta)$ of the boson model with Lennard-Jones
potential. There is a unique state (tangent functional) at high temperature, while a domain
with two extremal states, and hence long-range order (LRO), is present for low temperature and small hopping (darker zone).
Most of the phase diagram is not rigorously understood yet.}
\label{figphd}
\end{figure}

The proof of existence of phase transitions were obtained in \cite{BKU,DFF}; it was
realized in \cite{FRU} that tangent functionals naturally fit in the context of the
Pirogov-Sinai theory.

The zero-temperature energy takes the form (see \fig\ref{figgse})
\be
e^{\mu,h} = \min_{(\begin{smallmatrix}\cdot&\cdot\\
\cdot&\cdot\end{smallmatrix})} e^{\mu,h}(\begin{smallmatrix}\cdot&\cdot\\
\cdot&\cdot\end{smallmatrix})
\end{equation}
where the minimum is taken over the four configurations $(\begin{smallmatrix}0&0\\0&0\end{smallmatrix})$,
$(\begin{smallmatrix}1&0\\0&1\end{smallmatrix})$,
$(\begin{smallmatrix}0&1\\1&0\end{smallmatrix})$, and
$(\begin{smallmatrix}1&1\\1&1\end{smallmatrix})$. There are angles at the
intersections between different energies. It is not clear whether they subsist at finite
temperature however --- an example where angles disappear is the one-dimensional Ising
model. The main result of the Pirogov-Sinai theory, in this model, is the claim that there
exist four $C^1$
functions that are close to the energies \eqref{gsen}, and that play the same role: the free
energy is given by the minimum of these four functions, and hence has angles at their
intersections.

\begin{theorem}[Free energy at low temperature]
\label{thmbosmodel}

Assume $d \geq 2$.
Let $U(0) \to \infty$, $U(1)>0$ and $U(\sqrt2)<0$. There exist $\beta_0, r<\infty$
such that if $\beta \geq \beta_0$ and $t + u_r \leq 1$, there are real
functions $f^{\mu,h}_{(\begin{smallmatrix}0&0\\0&0\end{smallmatrix})}$,
$f^{\mu,h}_{(\begin{smallmatrix}1&0\\0&1\end{smallmatrix})}$,
$f^{\mu,h}_{(\begin{smallmatrix}0&1\\1&0\end{smallmatrix})}$,
$f^{\mu,h}_{(\begin{smallmatrix}1&1\\1&1\end{smallmatrix})}$ such that
\begin{itemize}
\item
$$
\limtwo{\beta\to\infty}{T,u_r \to 0} f^{\mu,h}_{(\begin{smallmatrix}\cdot&\cdot\\
\cdot&\cdot\end{smallmatrix})} = e^{\mu,h}(\begin{smallmatrix}\cdot&\cdot\\
\cdot&\cdot\end{smallmatrix}) \quad\quad \text{and} \quad \limtwo{\beta\to\infty}{T,u_r \to
0} \frac\partial{\partial \mu,h} f^{\mu,h}_{(\begin{smallmatrix}\cdot&\cdot\\
\cdot&\cdot\end{smallmatrix})} = \frac\partial{\partial \mu,h} e^{\mu,h}(\begin{smallmatrix}\cdot&\cdot\\
\cdot&\cdot\end{smallmatrix})
$$
uniformly in $\mu,h$. Limits are taken in any order. The limit $u_r\to0$ means that
$U(a)\to0$ for all $a\geq2$.
\item The free energy \eqref{deffen} is given by
$$
f^{\mu,h} = \min_{(\begin{smallmatrix}\cdot&\cdot\\
\cdot&\cdot\end{smallmatrix})} f^{\mu,h}_{(\begin{smallmatrix}\cdot&\cdot\\
\cdot&\cdot\end{smallmatrix})}.
$$
\item The functions are $C^1$ in $\mu,h$ with uniformly bounded derivatives. Furthermore, $f^{\mu,h}_{(\begin{smallmatrix}\cdot&\cdot\\
\cdot&\cdot\end{smallmatrix})}$ is real analytic in $\mu,h$ when $f^{\mu,h}_{(\begin{smallmatrix}\cdot&\cdot\\
\cdot&\cdot\end{smallmatrix})}$ is the unique minimum.
\end{itemize}
\end{theorem}

The phase diagram is therefore governed by these four functions; clearly, it is symmetric
under the transformation $h \to -h$. Let $\mu_1$ be the coexistence point of
$(\begin{smallmatrix}0&0\\0&0\end{smallmatrix})$ and the chessboards, \ie
\be
f^{\mu_1,0}_{(\begin{smallmatrix}0&0\\0&0\end{smallmatrix})} =
f^{\mu_1,0}_{(\begin{smallmatrix}0&1\\1&0\end{smallmatrix})},
\end{equation}
and $\mu_2$ be the coexistence between the chessboard and
$(\begin{smallmatrix}1&1\\1&1\end{smallmatrix})$.
There are exactly two extremal tangent functionals for
$\mu_1<\mu<\mu_2$ and $h=0$. Exactly three for $\mu=\mu_1$ and $h=0$, as well as for
$\mu=\mu_2$ and $h=0$. There is a unique tangent functional everywhere else.

Among the consequences are various first-order phase transitions. For instance,
\be
\frac\partial{\partial h} f^{\mu,h} \Big|_{h=0-} \neq \frac\partial{\partial h} f^{\mu,h}
\Big|_{h=0+}
\end{equation}
for $\mu_1<\mu<\mu_2$; also, if $h=0$,
\be
\frac\partial{\partial\mu} f^{\mu,0} \Big|_{\mu=\mu_1-} \neq \frac\partial{\partial\mu}
f^{\mu,0}
\Big|_{\mu=\mu_1+},
\end{equation}
and similarly at $\mu_2$.

Construction of the functions (`metastable free energies' in the Pirogov-Sinai
terminology) is done in two steps. First, using a space-time representation of the
model, one defines an equivalent {\it contour model}. This step is explained in the next
section; it gives the opportunity to make the link with a stochastic process of classical
particles
jumping on the lattice. The second step is to get an expression for the metastable free
energies starting from a contour model, and this is achieved using the standard Pirogov-Sinai
theory \cite{PS,Sin}. This is only outlined here. Ideas are described \eg in \cite{Kot}; we
also mention \cite{Uel} for a
self-contained review which includes precise statements on tangent functionals.

\subsection{Incompressibility}

The space-time contour representation actually allows us to obtain more. The total number
of particles is conserved, and as a consequence the ground state of the quantum model has
same density as that of the model without hoppings, and hence the compressibility is zero.
The following observations were made in \cite{BKU2}.

Since a state is a linear functional on the space of interactions, we have to understand
what is the {\it density} of the systems. We consider the interaction $N$:
\be
N_A = \begin{cases} \hat n_x & \text{if } A = \{x\} \\ 0 & \text{ otherwise;} \end{cases}
\end{equation}
if $\rho$ denotes a state, than the corresponding density is $\rho(N)$. It is a function of
the chemical potential $\mu$. One defines the {\it compressibility} $\kappa_{\text T}$,
\be
\kappa_{\text T} = \frac\partial{\partial\mu} \rho(N)
\end{equation}
where the derivative is with constant temperature (\ie $\beta$). The theorem below claims
incompressibility of the ground state, and also that the low temperature states are close
to incompressible. It holds in all dimensions.

\begin{theorem}
\label{thmincomp}

Let $U(0) \to \infty$, $U(1)>0$ and $U(\sqrt2)<0$. There exist $\beta_0,r<\infty$
such that if $\beta \geq \beta_0$ and $t + u_r \leq 1$, one has
\ba
&\bigl| \rho(N) - \rho_0(N) \bigr| \leq C \e{-\beta r'} ; \nn\\
&|\kappa_{\rm T}| \leq C \e{-\beta r'} \nn
\end{align}
for some $C<\infty$, $r'>0$.
\end{theorem}

\section{The space-time representation and the equivalent contour model}
\label{seccontrep}

\subsection{Equivalence with a stochastic system}

We start with the finite-volume expression for the free energy,
\be
\label{fenfin}
f_\Lambda^{\mu,h} = -\frac1{\beta|\Lambda|} \log\Tr \e{-\beta \sum_{A \subset \Lambda}
H_A},
\end{equation}
with $H=T+V-\mu N$. Notice that the last two
interactions are diagonal with respect to the basis \eqref{defbasis}.

One can give various probabilistic interpretations for \eqref{fenfin}, see \eg \cite{Toth}.
A natural one is a continous-time Markov chain where the collection of random variables $\{ n(t) \}_{t\geq 0}$ take values in
$\{0,\dots,N\}^\Lambda$. Let us introduce the set of `neighbors' of a configuration $n$:
\be
\caN(n) = \{ n': \exists x,y \text{ with } |x-y|=1 \text{ and } n_x' = n_z - \delta_{zx} +
\delta_{zy} \text{ for all } z \in \Lambda \}.
\end{equation}
The generator of this random process is
\be
G_{nn'} = \begin{cases} 1 & \text{if } n' \in \caN(n) \\ -|\caN(n)| & \text{if } n'=n \\ 0
& \text{otherwise.} \end{cases}
\end{equation}
The {\it partition function} $Z_\Lambda = \Tr \e{-\beta \sum_{A \subset \Lambda} H_A}$
is the expectation
\be
\label{partfctsp}
Z_\Lambda = \bbE_{[0,\beta]} \biggl( \chi[n(0)=n(\beta)] \exp\Bigl\{ -\int_0^\beta \dd\tau
\Bigl[ \sum_{x,y \in \Lambda} U(|x-y|) n_x(\tau) n_y(\tau) - \mu \sum_{x \in \Lambda} n_x(\tau) \Bigr] \Bigr\} \biggr).
\end{equation}

Another representation that is more appealing for the physical intuition involves
continu\-ous-time simple random walks. It was explicited in \cite{CS} and used to obtain a
bound on the free energy of the Heisenberg model \cite{CS2,Toth}. Let $\{ x_j(t) \}_{t\geq0}$, $1 \leq j \leq N$,
be random walks with generator
\be
L_{xy} = \begin{cases} 1 & \text{if } |x-y|=1 \\ -2d & \text{if } x=y \\ 0 &
\text{otherwise.} \end{cases}
\end{equation}
Then the partition function takes the form
\bm
\label{partfctrw}
Z_\Lambda = \sum_{N\geq0}^\infty \frac{\e{\beta\mu N}}{N!} \sum_{x_1,
\dots,x_N \in \Lambda} \sum_{\pi \in S_N} \bbE\biggl( \chi\bigl[ x_i(\beta) = x_{\pi(i)}, 1\leq i \leq N
\bigr] \\
\exp\Bigl\{ -\int_0^\beta \dd\tau \sum_{i<j} U(|x_i(\tau)-x_j(\tau)|) \Bigr\} \bigg|
\, x_i(0) = x_i, 1\leq i\leq N \biggr).
\end{multline}
Here particles have to start and end in $\Lambda$, but they are meanwhile free to move outside.
One could impose more stringent boundary conditions, by defining a generator
$L^\Lambda_{xy}$ that does not allow particles to leave or enter $\Lambda$, or by adding an
infinite potential outside of $\Lambda$. It is however useless, as the free
energies corresponding to these various partition functions have the same thermodynamic limit.

Notice the sum over permutations in \eqref{partfctrw}; this suggests
to consider probability on sets of permutations, for instance the probability that the
permutation has infinite cycles. We discuss this in Section \ref{secinfcyc},
where \eqref{partfctrw} is heuristically important.

Let us mention another example of close ties between quantum systems and probability
theory: Aizenman and Nachtergaele showed the equivalence of a quantum spin chain with a
stochastic process, which is itself equivalent to a two-dimensional Potts model
\cite{AN}. Using results established for the latter, the authors can draw new conclusions on the former.

\subsection{Equivalence with a contour model}

A way to derive these stochastic representations is by using Duhamel formula: if $\bsA$ and $\bsB$ are two
matrices, then
\ba
\e{\bsA+\bsB} &= \e\bsA + \int_0^1 \dd\tau \e{\tau\bsA} \bsB \e{(1-\tau) (\bsA+\bsB)} \nn\\
&= \e\bsA + \sum_{m\geq1} \int_{0<\tau_1<...<\tau_m<1} \dd\tau_1 \dots \dd\tau_m \e{\tau_1
\bsA} \bsB \e{(\tau_2-\tau_1) \bsA} \bsB \dots \bsB \e{(1-\tau_m) \bsA}.
\label{Duhamel}
\end{align}

Here we set $\bsA = \sum_{A \subset \Lambda} (V_A - \mu N_A - h P_A)$, with $P$ denoting
the staggered interaction, and $\bsB = \sum_{A
\subset \Lambda} T_A$. Taking the trace, and introducing $\bbbone = \sum_n \ket n \bra n$
on the right of each operator $\bsB$, we get the following expression:
\bm
\label{sptirep}
Z_\Lambda = \sum_{m\geq0} (-1)^m \sum_{A_1,\dots,A_m} \sum_{n_1,\dots,n_m}
\int_{0<\tau_1<...<\tau_m<\beta} \dd\tau_1\dots\dd\tau_m \e{-\tau_1 H_\Lambda(n_1)}
\bra{n_1} T_{A_1} \ket{n_2} \\
\e{-(\tau_2-\tau_1) H_\Lambda(n_2)} \bra{n_2} T_{A_2} \ket{n_3}
\dots \bra{n_m} T_{A_m} \ket{n_1} \e{-(\beta-\tau_m) H_\Lambda(n_1)},
\end{multline}
where we introduced $H_\Lambda(n) = \sum_{A\subset\Lambda} \bra n H_A \ket n$. One
recognizes \eqref{partfctsp} and \eqref{partfctrw}. Indeed, the sum over $\{A_i\}$ is
actually over pairs of nearest-neighbors; $\bra{n_1} T_{\{x,y\}} \ket{n_2}$ is zero unless
$n_2$ is a `neighbor' of $n_1$, \ie it is the same as $n_1$ up to one particle that moved
from $x$ to $y$, or from $y$ to $x$. Finally, $\{ \e{-(\tau_j-\tau_{j-1}) H_\Lambda(n_j)}
\}$ is represented in \eqref{partfctsp} and \eqref{partfctrw} by the exponential.

To each choice of $m$, $\{A_j\}$, $\{n_j\}$, $\{\tau_j\}$, corresponds a space-time
picture illustrated in \fig\ref{figsptime}. We write $\bsn(\tau)$ the configuration at time
$\tau$, that is, $\bsn(\tau) = n_j$ if $\tau_{j-1} \leq \tau < \tau_j$.

\bfig
\epsfxsize=80mm
\centerline{\epsffile{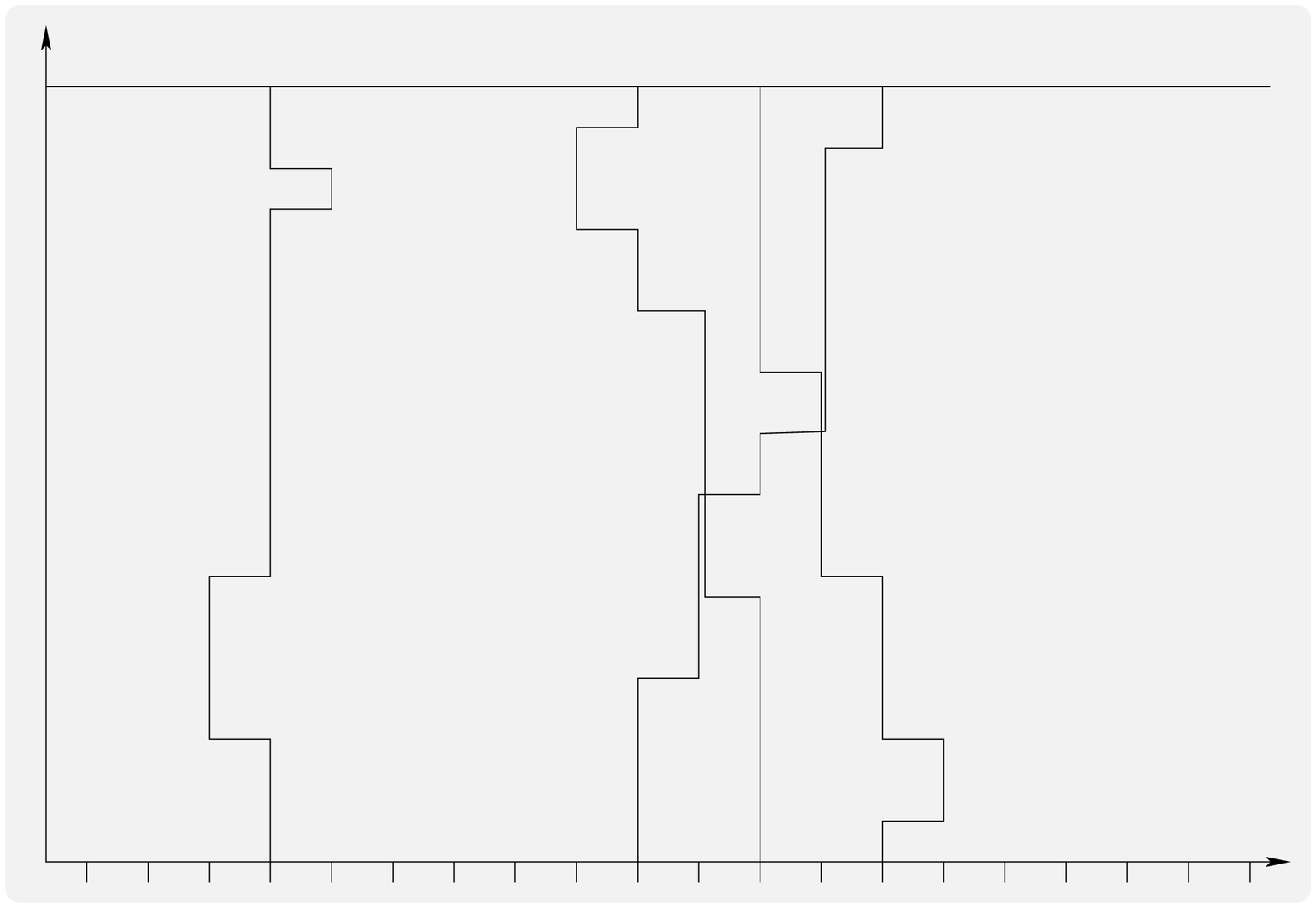}}
\figtext{
\writefig   3.8  0.6  {\small $\Lambda$}
\writefig   -4.05  0.75  {\footnotesize 0}
\writefig   -4  5.3  {\small $\beta$}
}
\caption{Space-time representation of the partition function as expanded in
\eqref{sptirep}. There are four bosonic trajectories in this picture.}
\label{figsptime}
\end{figure}

The goal is to extract some information on the analytic properties of the free energy, that
is, the logarithm of the partition function. A technique that was proposed in 1975 for the
study of extensions of the Ising model is the Pirogov-Sinai theory \cite{PS,Sin}, which was
later extended to quantum systems in \cite{BKU,DFF,DFFR,KU,FRU}. The strategy is to map the
quantum system onto
a `contour model'. The latter is a model where the states are not configurations or vectors of a
Hilbert space, but sets of mutually disjoint contours; the statistical weight $\e{-\beta
H}$ is replaced by a product of individual weights for each contour.

Let us describe in details the setting of a contour model.

A {\it contour} $\caA$ is a pair $(A,\alpha)$, where $A \subset
\bbZ^\nu$ is a connected set and is the {\it support} of $\caA$. In order to define
$\alpha$, let us introduce the closed unit cell $C(x) \subset
\bbR^\nu$ centered at $x$; the {\it boundary} $B(A)$ of $A \subset \bbZ^\nu$ is
\be
B(A) = \{ C(x) \cap C(y) : x \in A, y \notin A \}.
\end{equation}
The boundary $B(A)$ decomposes into connected components; each connected 
component $b$ is
given a label $\alpha_b \in \{1,\dots,p\}$, and $\alpha = (\alpha_b)$.

Let $\Lambda \subset \bbZ^\nu$ finite, with periodic boundary
conditions. A set of contours $\{\caA_1, \dots, \caA_k\}$ is {\it
admissible} iff
\begin{itemize}
\item $A_i \subset \Lambda$ for all $i$, and $\dist(A_i,A_j) \geq 1$ if $i\neq j$.
\item Labels $\alpha_j$ are matching in the following sense. Let $W = \Lambda 
\setminus
\cup_{j=1}^k A_j$; then each
connected component of $W$ must have same label on its boundaries.
\end{itemize}
For $j \in \{1,\dots,p\}$, let $W_j$ be the union of all connected
components of $W$ with labels $j$ on their boundaries.

The partition function of a contour model has the form
\be
Z_\Lambda = \sum_{\{\caA_1,\dots,\caA_k\}} \prod_{j=1}^k  w(\caA_j)
\prod_{i=1}^p \e{-\beta e_i |W_i|},
\end{equation}
where the sum is over admissible sets of contours in $\Lambda$.

The {\it weight} $w(\caA)$ of a contour $\caA$ is a complex 
function of the temperature and of the parameters of the phase diagram (here $\mu$ and $h$)
that is real anlaytic in all these parameters. Furthermore, we need that
\be
\label{boundweight}
|w(\caA)| \leq \e{-\beta e_0 |A|} \e{-r|A|}
\end{equation}
for a large enough constant $r$ (depending on $d$ and $p$). This typically holds when
$\beta$ is large. We also need that
partial derivatives of the weights with respect to $\mu$ and $h$ satisfy the same
bounds.

Many classical lattice models have such a representation. The usual way to define a contour
model is to attribute a set of contours to each configuration. One is given a finite set of
periodic configurations (`low energy configurations', or `reference configurations'), and
one defines `excited sites' as
those sites whose neighborhood does not agree with any of the reference configurations. The
set of excited sites decompose into connected components, that are supports of the
contours. Outside the contours the configuration agrees with one of the reference configurations,
and the labels indicate which one.

The labels are important because the weight of a contour typically depends on which
configuration lies outside. If we want this weight to depend on the contour only, we need
to provide the information contained in the labels.

We are looking for a similar approach here with the space-time representation. On the one
hand, we expect the phase diagram to display four phases: a phase with very low density,
corresponding to $(\begin{smallmatrix}0&0\\0&0\end{smallmatrix})$; two chessboard phases,
$(\begin{smallmatrix}1&0\\0&1\end{smallmatrix})$ and
$(\begin{smallmatrix}0&1\\1&0\end{smallmatrix})$; and a phase with density close to 1,
$(\begin{smallmatrix}1&1\\1&1\end{smallmatrix})$. These are our reference configurations.
On the other hand, we suppose here that
particles have small hoppings, so that jumps are typically rare in \fig\ref{figsptime}.

In order to get contours that have supports on a lattice, we discretize the continuous
direction. Let $\tilde\beta$ such that $\beta = M \tilde\beta$ with $M$ an integer. We
consider the lattice $\bsLambda = \Lambda \times \{1,\dots,M\}$. A site $\bsx = (x,s) \in
\bsLambda$ is `in the state $(\begin{smallmatrix}0&0\\0&0\end{smallmatrix})$' if for all
$y$ with $|y-x|\leq1$, and all $(s-1)\tilde\beta < \tau < s\tilde\beta$, we have
$\bsn_y(\tau) = 0$. We make similar definitions for the other three reference
configurations.

Cells that are not in such a state are {\it excited}. Connected components of the set of
excited cells are the supports of the contours, and labels take values in $\{
(\begin{smallmatrix}0&0\\0&0\end{smallmatrix}),
(\begin{smallmatrix}1&0\\0&1\end{smallmatrix}),
(\begin{smallmatrix}0&1\\1&0\end{smallmatrix}),
(\begin{smallmatrix}1&1\\1&1\end{smallmatrix}) \}$ and contain information on which
configuration touches the support. This is illustrated in \fig\ref{figcont}.

\bfig
\epsfxsize=80mm
\centerline{\epsffile{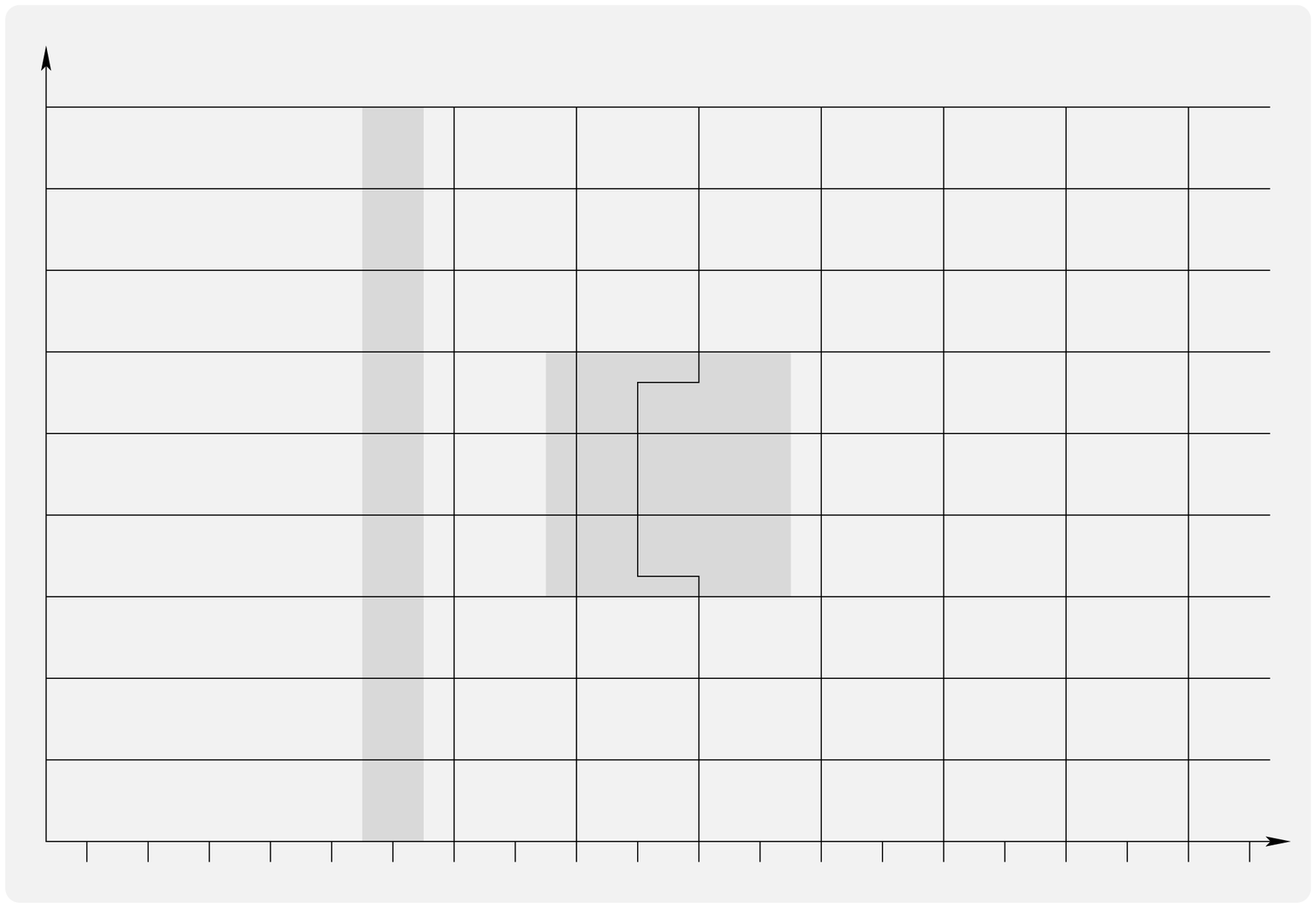}}
\figtext{
\writefig  3.8  0.7  {\small $\Lambda$}
\writefig   -4.05  0.85  {\footnotesize 0}
\writefig   -4  5.2  {\small $\beta$}
}
\caption{Contours in the space-time representation. The contour on the left separates the
empty configuration from a chessboard one, while the one on the right is due to the motion
of a particle.}
\label{figcont}
\end{figure}

Summing first over contour configurations, then integrating over compatible space-time
configurations, we can rewrite \eqref{sptirep} as
\be
Z_\Lambda = \sum_{\{\caA_1,\dots,\caA_k\}} \prod_{i=1}^k w(\caA_i) \prod_{i=1}^4
\e{-\tilde\beta e_i |W_i|}.
\end{equation}
The expression for the weight $w(\caA)$ is complicated, but the exponential bound
\eqref{boundweight} is not too hard to obtain. It will require $\tilde\beta\Delta$ to be
large, and $\tilde\beta t$ to be small. Theorem \ref{thmbosmodel} is then a result of the
Pirogov-Sinai theory, see for instance \cite{Uel}.

\subsection{Consequences of the contour representation}

A few words need to be added in view of Theorem \ref{thmincomp}. The density is
\be
\label{rho}
\rho = \lim_{\Lambda \nearrow \bbZ^d} \frac{\Tr\bigl( \frac1{|\Lambda|} \sum_{x \in
\Lambda} \hat n_x \bigr) \e{-\beta \sum_{A\subset\Lambda} H_A}}{Z_\Lambda}.
\end{equation}
This expression for the density agrees with that in terms of derivative of the free energy,
provided the latter is differentiable. Indeed, let $f(\mu)$ be the infinite volume free
energy as a function of the chemical potential. It is concave, and if it is differentiable at
$\mu$ we have
\be
\rho = -\frac{\dd f(\mu)}{\dd\mu} = -\lim_{\Lambda \nearrow \bbZ^d} \frac{\dd
f_\Lambda(\mu)}{\dd\mu}.
\end{equation}

The space-time expansion of \eqref{rho} was studied in \cite{BKU2}. Due to the conservation
of the total number of particles, differences between the density of the quantum model
(with hoppings) and the classical one (without hoppings) lead to contours that wind around
the torus $\Lambda \times [0,\beta]_{\text{per}}$. Hence their length is at least $\beta$, and no
such contours survive when taking the limit $\beta\to\infty$. As a consequence, the density
of the quantum model is locked to the classical one.

This clearly implies that the compressibility vanishes at zero temperature. To obtain the
low temperature bounds requires some more work, that also goes through an expansion involving
winding contours \cite{BKU2}.

\subsection{Proof of Theorem \ref{thmht}}
\label{subsecproofht}

We conclude this section by proving that there is a unique equilibrium state at high
temperature, as stated in Theorem \ref{thmht}. It strongly relies on ideas discussed above,
with many simplifications. We show the equivalence between the quantum model and a {\it
polymer model} --- this is a contour model without labels (\ie $p=1$). Once we have
obtained this equivalence, the results follow from cluster expansions \cite{KP,Dob,BZ}.

Using the Duhamel formula \eqref{Duhamel}, we get
\bm
\Tr \e{-\beta \sum_{A \subset \Lambda} H_A} = \Tr \e{-\beta \sum_{x \in \Lambda} H_{\{x\}}}
+ \sum_{m\geq1} (-1)^m \sumtwo{A_1,\dots,A_m}{|A_j|\geq2} \int_{0<\tau_1<...<\tau_m<\beta}
\dd\tau_1 \dots \dd\tau_m \\
\Tr \e{-\tau_1 \sum_x H_{\{x\}}} H_{A_1} \e{-(\tau_2-\tau_1)
\sum_x H_{\{x\}}} \dots \e{-(\beta-\tau_m) \sum_x H_{\{x\}}}.
\end{multline}
Let $F_x^{(0)}$ be defined by
\be
\e{-\beta F_x^{(0)}} = \Tr \e{-\beta H_{\{x\}}},
\end{equation}
with the trace taken in the single-site Hilbert space $\caH_x$. We also set $F_A^{(0)} =
\sum_{x\in A} F_x^{(0)}$. We define polymers as
connected components of the set $\cup_{j=1}^m A_j$, and the weight of a polymer $A$ to be
\bm
\label{weightpol}
w(A) = \e{\beta F_A^{(0)}} \sum_{m\geq1} (-1)^m \sumtwo{A_1,\dots,A_m}{|A_j|\geq2, \,\cup_j A_j
= A} \int_{0<\tau_1<...<\tau_m<\beta}
\dd\tau_1 \dots \dd\tau_m \\
\Tr \e{-\tau_1 \sum_{x\in A} H_{\{x\}}} H_{A_1} \e{-(\tau_2-\tau_1)
\sum_x H_{\{x\}}} \dots \e{-(\beta-\tau_m) \sum_{x\in A} H_{\{x\}}}.
\end{multline}
Then
\be
\Tr \e{-\beta \sum_{A \subset \Lambda} H_A} = \e{-\beta F_\Lambda^{(0)}}
\sum_{\{A_1,\dots,A_k\}} \prod_{j=1}^k w(A_j).
\end{equation}
This is the partition function of a polymer model.

We need a bound on the weight of the polymers. Since the dimension of the Hilbert space
$\caH_A$ is $(N+1)^{|A|}$, we can estimate the last line of \eqref{weightpol} by
\be
\Tr\cdot \leq (N+1)^{|A|} \prod_{x\in A} \| \e{-\beta H_{\{x\}}} \| \prod_{j=1}^m \|H_{A_j}\|.
\end{equation}
Furthermore,
\be
\| \e{-\beta H_{\{x\}}} \| \leq \e{-\beta F_x^{(0)}},
\end{equation}
so we obtain
\ba
|w(A)| &\leq (N+1)^{|A|} \e{-r|A|} \sum_{m\geq 1} \frac1{m!} \Bigl( \sumtwo{A'\subset
A}{|A'|\geq2} \|H_{A'}\| \e{r|A'|} \Bigr)^m \nn\\
&\leq \e{|A| \log(N+1)} \e{-r|A|} \e{|A| \|H\|_r^*}.
\end{align}

This satisfies the assumptions of the cluster expansions when $\|H\|_r^* \leq 1$ and
$r-\log(N+1)-1$ is large enough (depending on $d$ only). One then obtains an exact
expression for the infinite-volume free energy: in the translation invariant case
($F_x=F_y$ and $w(A+x)=w(A)$), the mean free energy is given by
\be
\label{fclusters}
f = F_0^{(0)} - \frac1\beta \sum_{(A_1,\dots,A_k)} \varphi^{\text T}(A_1,\dots,A_k) \prod_{j=1}^k w(A_j),
\end{equation}
with the sum over {\it clusters}, that is, $k$-tuples $(A_1,\dots,A_k)$, $k\geq1$, such that their
union $\cup_{j=1}^k A_j$ is connected. The combinatoric factor $\varphi^{\text T}(A_1,\dots,A_k)$ has an
expression involving the
graph of $k$ vertices with an edge between $i$ and $j$ whenever $A_i \cup A_j$ is
connected. The results on cluster expansions include bounds ensuring the convergence of the
sum \eqref{fclusters}; see \eg \cite{KP,Dob,BZ} for detailed results and proofs.

By averaging over a cell whose dimensions are given by the periods of the interactions, one
obtains a similar expression in the case of periodicity rather than translation invariance.

If $\|H\|_r^* < 1$, then $\|H+\lambda P\|_r^* < 1$ for all perturbation $P$, and $\lambda$
in a neighborhood of 0, and one can perform the above expansions. As a result, we obtain a
free energy $f(\lambda)$ that is given by a convergent sum of clusters, with weights that
are
analytic in $\lambda$. Therefore $f(\lambda)$ is real analytic, and there is a unique
tangent functional at $H$.

\section{A discussion of the Bose-Einstein condensation}
\label{secBEC}

\subsection{The origins}

The story started in 1924 when Bose sent a paper to Einstein, that was previously rejected
by {\it Philosophical Magazine}. Einstein translated it into German and recommended its
publication in {\it Zeitschrift f\"ur Physik}; he wrote articles shortly afterwards in {\it
Sitzungsberichte der Preussische Akademie der Wissenschaften} (1924--25). The
`Bose-Einstein statistics' for quantum particles (in particular photons) was uncovered,
and a curious phase transition was proposed, where the ground state of the one-particle
Hamiltonian is macroscopically occupied. This is the {\it Bose-Einstein condensation} for
the {\it ideal} boson gas (that is, without interactions).

For some time it was not clear whether such a transition was really occurring in the
nature; but London proposed in 1938 that superfluidity in Helium was a
consequence of a Bose-Einstein condensation, an idea that is largely accepted nowadays.

Is there a condensation for interacting systems as well, and what does it mean? These
questions were addressed by Feynman
\cite{Fey}; he proposed the idea that the transition corresponds to positive probability
for the occurrence of
infinite cycles in the space-time representation --- this will be discussed in greater
details in the next section. Feynman's conclusion is that weakly interacting systems behave like
non-interacting ones, albeit with a larger effective mass, and still display condensation.

Direct experimental evidence of BEC has been observed only recently \cite{AEMWC}.

\subsection{General ideas}

A system of $N$ bosons in the continuum is described by the Hamiltonian
\be
\label{genHam}
H = -\frac{\hbar^2}{2m} \sum_{j=1}^N \Delta_j + \sum_{i<j} U(|x_i-x_j|).
\end{equation}
Here, $\Delta_j$ is the Laplace operator $\sum_{\alpha=1}^d \frac{\partial^2}{\partial
x_{j,\alpha}^2}$ and $x_j$ is the position of the $j$-th particle. 
The low temperatures should be described by the Bogolubov theory, see \eg \cite{Lieb, ZB}
for an introduction and partial justifications. The Bogolubov theory relies on the
assumption that most of the particles are in the ground state of the Laplace operator (that
is, the Hamiltonian for the ideal gas), which is false in presence of interactions. Still,
many predictions are correct; in particular, it gives a value for the ground state
energy per particle $e_0$ at low density,
\be
e_0 = \frac{2\pi\hbar^2\rho a}m \bigl( 1 + o(\rho a^3) \bigr),
\end{equation}
where $\rho$ is the density and $a$ is the scattering length of the potential $U$.
This formula has been rigorously established by Lieb and Yngvason \cite{LY}. This
and other results are reviewed in \cite{Lieb2}.

Further developments led to the concept of {\it off-diagonal long-range order} due to
Penrose and Onsager \cite{PO}.
Take \eg the lattice model of Section \ref{secexample}. One considers the following order
parameter:
\be
\expval{c_x^\dagger c_y} = \lim_{\Lambda \nearrow \bbZ^d} \frac{\Tr c_x^\dagger c_y
\e{-\beta \sum_{A \subset \Lambda} H_A}}{\Tr \e{-\beta \sum_{A \subset \Lambda} H_A}}
\end{equation}
where $H = T + V - \mu N$, and the traces are in the Hilbert space $\otimes_{x\in\Lambda}
\caH_x$. Here, it is natural to set periodic boundary conditions for $\Lambda$. The
question is:
\begin{quote}
{\it Does $\lim_{|x-y|\to\infty} \expval{c_x^\dagger c_y}$ differ from 0?}
\end{quote}

The equilibrium state at high temperature is unique and clustering, see Theorem
\ref{thmht}, and hence BEC must be searched at low temperatures.

\subsection{The hard-core boson lattice model}

There is one rigorous result concerning the existence of condensation in a reasonable model
of interacting bosons. This is a lattice model where
bosons interact with hard-core repulsion, \ie the Hamiltonian \eqref{defHam} with
$U(0)\to\infty$ and $U(a)\to0$ if $a\geq1$. The theorem below is due to Dyson, Lieb and
Simon \cite{DLS}, and Kennedy, Lieb and Shastry \cite{KLS}. It is stated for 3 or more
dimensions and at low temperature, but it also holds for the ground state of the
2-dimensional model \cite{KLS}.

\begin{theorem}
\label{thmBEChcb}
Take $d\geq3$, $H = T + V$ with $U(0) \to \infty$, $U(a) \to 0$ for $a>0$. Then there is $\beta_0 < \infty$ such that for $\beta>\beta_0$,
$$
\lim_{|x-y|\to\infty} \expval{c_x^\dagger c_y} \neq 0.
$$
\end{theorem}

This theorem implies the existence of a phase transition in the sense that the state
$\expval\cdot$ is not clustering. It is established using `reflection positivity', introduced in \cite{FSS} for
proving spontaneous magnetization in the classical Heisenberg model; its difficult
extension to quantum systems was done in \cite{DLS}. The claims of \cite{DLS,KLS} that are
relevant here deal with spontaneous magnetization in the spin $\frac12$ $x$-$y$ model. Let
us discuss analogies between spins and hard-core boson systems. For the latter, we take $\caH_0 \simeq \bbC^2$ and define self-adjoint operators
$\{S_x^{(1)}, S_x^{(2)}, S_x^{(3)}\}_{x\in \bbZ^d}$, that commute if they are located on
different sites, and satisfy $[S_x^{(1)}, S_x^{(2)}] = \ii S_x^{(3)}$ (and permutations of (1,2,3)) at a
same site. (These matrices are called {\it Pauli matrices}.) The $x$-$y$ model has
interaction $-S_x^{(1)} S_y^{(1)} - S_x^{(2)} S_y^{(2)}$ on nearest-neighbor sites $x,y$,
and zero otherwise.

The correspondence to boson models is done by setting
\ba
&c_x^\dagger = S_x^{(1)} + \ii S_x^{(2)} \nn\\
&c_x = S_x^{(1)} - \ii S_x^{(2)} \\
&n_x = S_x^{(3)} + \tfrac12 \nn
\end{align}
In the case of hard-cores (with $N=1$) the commutation relations are $[c^{\#}_x, c_y^{\#}] = 0$ if $x
\neq y$, and $\{ c_x, c_x^\dagger \} = 1$, where $\{\cdot,\cdot\}$ denotes the
anticommutator. It is easy to check that these also follow from
the commutation relations of spin operators, and from definitions above. The
$x$-$y$ model is equivalent to $H'$,
\be
H_A' = \begin{cases} -\tfrac12 [c^\dagger_x c_y + c^\dagger_y c_x] & \text{if } A
= \{x,y\}, \quad |x-y|=1 \\ 0 & \text{otherwise.} \end{cases}
\end{equation}
Off-diagonal long-range order is then equivalent to spontaneous magnetization in the 1-2
plane.

\subsection{BEC \& symmetry breaking}

The Bose-Einstein condensation is related to a symmetry breaking, namely `global gauge
invariance'. Let us note that the Hamiltonian \eqref{defHam} conserves the
total number of particles, \ie
\be
\bigl[ \sum_{A \subset \Lambda} H_A, \sum_{x \in \Lambda} \hat
n_x \bigr] = 0.
\end{equation}
Therefore one can define the unitary operator $U_\Lambda = \e{\ii \alpha \sum_{x \in
\Lambda} \hat n_x}$, which is a symmetry of the Hamiltonian. Its action on creation and
annihilation operators is
\ba
& U_\Lambda c_x^\dagger U_\Lambda^{-1} = \e{\ii\alpha} c_x^\dagger \nn\\
& U_\Lambda c_x U_\Lambda^{-1} = \e{-\ii\alpha} c_x^\dagger.
\end{align}
This is easily seen from the action of all these operators on elements of the basis
\eqref{defbasis}.

To study the properties of the free energies as a function of the interactions, one has to
proceed similarly as in Section \ref{secexample}. Recall that we added a non
translation-invariant (and non-physical) interaction $h P$ and looked at a phase diagram
where $h$ is a parameter. This is similar here. First, we need an interaction that does not
conserve the total number of particles. The simplest choice with self-adjoint operators is
$Q = (Q_A)$, with
\be
Q_A = \begin{cases} \e{\ii\alpha} c_x^\dagger + \e{-\ii\alpha} c_x & \text{if } A = \{x\}
\\ 0 & \text{otherwise.} \end{cases}
\end{equation}
Supposedly, there is a unique tangent functional to the free energy at $H + h Q$ for all
$h\neq0$, but there should be an infinite number of extremal states at $H$, if the
temperature is low enough; each of these extremal states is indexed by $\alpha \in
[0,2\pi)$. Since there is a unique equilibrium state at high temperature (Theorem
\ref{thmht}), we face here the
breakdown of a continuous symmetry. It should occur at low temperature and if the dimension
of the lattice is greater or equal to 3.

There is no rigorous result to support this discussion, besides the weaker --- but
important! --- statement of Theorem
\ref{thmBEChcb} in the case of the hard-core boson gas.

\section{Infinite cycles: context and conjectures}
\label{secinfcyc}

\subsection{Heuristics}

In the last section of this brief review, we discuss an approach to the BEC initiated by
Feynman 50 years ago \cite{Fey}, that focusses on the occurrence of infinite cycles in the
space-time representation. Its appeal to probabilists should be evident --- it looks at
first sight like a percolation phenomenon. However, the one-dimensional nature of cycles
makes them harder to study than clusters. Still, some progress should be possible.

The partition function for the Hamiltonian \eqref{genHam} can be expanded via Feynman-Kac;
setting $\hbar^2/2m=1$, the partition function is given by
\bm
\label{fpartcontinuum}
Z_V = \sum_{N\geq0} \frac{\e{\beta\mu N}}{N!} \int_V \dd x_1 \dots \dd x_N \sum_{\pi \in
S_N} \Bigl( \prod_{i=1}^N \inttwo{\bsx_i(0)=x_i}{\bsx_i(\beta)=x_{\pi(i)}} \dd
W_{[0,\beta]}(\bsx_i) \Bigr) \\
\exp\Bigl\{ -\int_0^\beta \dd\tau \sum_{i<j}
U(|\bsx_i(\tau)-\bsx_j(\tau)|) \Bigr\}.
\end{multline}
Here, integrals are over Brownian paths starting at $x_i$ and ending at $x_{\pi(i)}$. See
\cite{Gin} for an introduction to functional integration. This expression is very similar
to \eqref{partfctrw} for lattice systems and is illustrated in \fig\ref{figfeykac}.

\bfig
\epsfxsize=80mm
\centerline{\epsffile{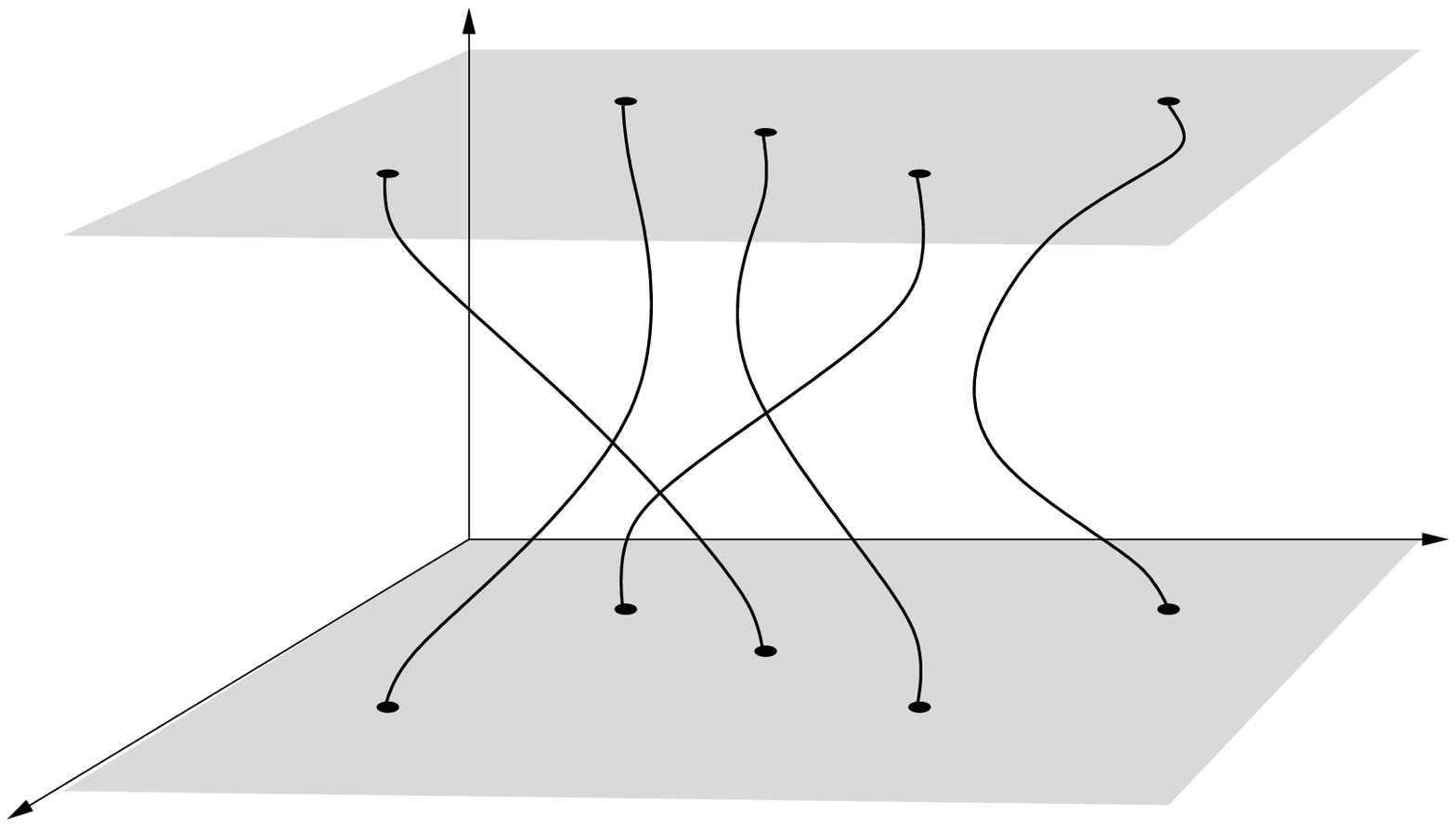}}
\figtext{
\writefig   -4.1  0.4  {\small $x$}
\writefig   3.9 2.0  {\small $y$}
\writefig   -1.8  4.7  {\footnotesize $\beta$}
\writefig   -1.9  0.95  {\tiny 1}
\writefig   -0.85  4.55  {\tiny $\pi(1)$}
}
\caption{Feynman-Kac representation of the partition function for bosons in the continuum.
The picture shows a situations with five particles and two cycles, of lengths 4 and 1.}
\label{figfeykac}
\end{figure}

The space-time is periodic in the vertical direction, so it is topologically equivalent to
a cylinder. Bosons wind around the cylinder, forming cycles (see \fig\ref{figfeykac}). Feynman's idea is to consider the length of these cycles, and to look at the
probability of occurrence of infinite ones. He identifies the onset of a positive
probability to a Bose-Einstein condensation. In his paper \cite{Fey} he argues that
interactions only slow down the diffusion of bosons, without forbidding infinite cycles, and
he concludes that BEC should also occur in interacting systems.

Cycles were studied in \cite{Suto}, where it is proved in particular that, in the case of the
ideal gas (that is, non-interacting particles), infinite cycles do occur below the transition
temperature for BEC. The converse statement, namely absence of infinite cycles in absence
of BEC, is not proven yet, although it is doubtlessly true.

However, the equivalence between BEC and occurrence of infinite cycles is not obvious. Consider
\eg the model discussed in Section \ref{secexample}. Our results imply absence of BEC at
low temperature and with small $T$; on the other hand, even though they have restricted
motions, bosons can interchange with neighbors, and infinite
cycles seem likely for low enough temperature, if the dimension is greater or equal to 3
--- this has something to do with probabilities of recurrence of random walks. A
lattice model can be viewed as a continuum model where the particles have condensed (in the
usual sense) and are displaying long-range order. The following conjecture is compatible
with these considerations:

\begin{quote}
{\bf Conjecture.}
{\it
\begin{itemize}
\item Occurrence of BEC implies positive probability of infinite cycles.
\item Positive probability of infinite cycles, and absence of long-range order, imply
occurrence of BEC.
\end{itemize}
}
\end{quote}

In the hope of shedding some light on this discussion, we introduce a simple lattice model
of cycles, state some (rather obvious) properties and propose some conjectures.

\subsection{A simple lattice cycles model}

The expression \eqref{fpartcontinuum} for the partition function starts by an integration over all initial
positions of the particles; let us suppose that they are located on the sites of the lattice
$\bbZ^d$ --- assuming that density fluctuations do not play an important role in the onset
of BEC, this assumption is a mild one at low temperature. Furthermore, we replace the integral over Brownian paths by an effective weight
$$
\prod_{x \in \Lambda} \e{-\xi(x,\pi(x))}
$$
(with $\Lambda \subset \bbZ^d$ finite). A natural choice for $\xi(x,y)$ is $|x-y|^2/\beta$,
with $|\cdot|$ the Euclidian distance, and $\beta$ the inverse temperature. Indeed, the Brownian paths for a time interval
$[0,\beta]$ diffuse like $\sqrt\beta$. Other choices are possible, for instance
$|x-y|^\gamma/\beta$ with $\gamma>2$ to account for large interactions. One could also
simplify the problem and consider
\be
\label{xinn}
\xi(x,y) = \begin{cases} 0 & \text{if } x=y \\ 1/\beta & \text{if } |x-y|=1 \\ \infty &
\text{otherwise.} \end{cases}
\end{equation}

In any case, we restrict the choice of $\xi$ to one that satisfies
\be
\sum_x \e{-\xi(0,x)} < \infty,
\end{equation}
ensuring that particles do not jump to infinity in one step.

Let us describe carefully these cycles models.

The lattice is $\bbZ^d$, and we denote by $\bbB$ the set of bijections $\bbZ^d \to \bbZ^d$.
Given $x,y \in \bbZ^d$, let $B_{xy} = \{ \pi \in \bbB : \pi(x)=y \}$; then we define
$\caB'$ to be the algebra made out of all such sets and their complements.

Next we set $\bbB(\Lambda) = \{ \pi \in \bbB : \pi(x)=x \text{ for all } x \notin \Lambda
\}$ the set of permutations that are trivial out of $\Lambda$. Since $\caB'$ is countable,
there exists a sequence of boxes $\bsLambda = (\Lambda_n)_{n\geq0}$ such that for all $B
\in \caB'$ the following limit exists:
\be
\label{defprob}
\lim_{\Lambda \in \bsLambda} \frac1{Z(\Lambda)} \sum_{\pi \in \bbB(\Lambda)}
\indicator{\pi \in B} \prod_{x\in\Lambda} \e{-\xi(x,\pi(x))} \equiv P(B).
\end{equation}
The normalization $Z(\Lambda)$ is
\be
Z(\Lambda) = \sum_{\pi \in \bbB(\Lambda)} \prod_{x\in\Lambda} \e{-\xi(x,\pi(x))}.
\end{equation}
The probability \eqref{defprob} extends to the smallest $\sigma$-algebra generated by
$\caB'$, that we denote $\caB$.

A {\it cycle} is a sequence $c = (x_1, \dots, x_{|c|})$ of different sites; we identify
$(x_2, \dots, x_{|c|}, x_1) = (x_1, \dots, x_{|c|})$. The set of permutations $B_c = \{ \pi
\in \bbB : \pi(x_j) = x_{j+1}, 1\leq j\leq n \}$ (with $x_{|c|+1} \equiv x_1$) is an element
of $\caB$, and the set of cycles is countable. Therefore, the set
\be
B_\infty = \bbB \setminus \union_{c \ni 0} B_c
\end{equation}
is also in the $\sigma$-algebra $\caB$. It represents the event `the origin belongs to an infinite
cycle', and is the central object of our attention.

\subsection{Few results and important conjectures}

There are no infinite cycles at high temperature; the condition of the following theorem is easy to
check for small $\beta$.

\begin{theorem}
If
$$
\sum_{c \ni 0} \prod_{j=1}^{|c|} \e{-\xi(x_j,x_{j+1})} < \infty,
$$
then $P(B_\infty)=0$.
\end{theorem}

\begin{proof}
Let $B_{>n}$ be the set of permutations where the origin belongs to a cycle of
length greater than $n$. One has
\be
B_{>1} \supset B_{>2} \supset \dots \quad \text{and} \quad B_\infty = \inter_n
B_{>n}.
\end{equation}
Then $P(B_\infty) = \lim_n P(B_{>n})$. Since
\be
B_{>n} = \union_{x\neq0} \, \uniontwo{w:0\to x}{|w|=n} B_w,
\end{equation}
with $w=(0,x_1,\dots,x_{n-1},x)$ a self-avoiding walk from 0 to $x$, and $B_w = \cap_{j=1}^n
B_{x_{j-1},x_j}$, one can write
\ba
P(B_{>n}) &= \sum_{x\neq0} \sumtwo{w:0\to x}{|w|=n} \lim_{\Lambda \in \bsLambda}
P_\Lambda(B_w) \nn\\
&\leq \lim_{\Lambda \in \bsLambda} \sum_{x\neq0} \sumtwo{w:0\to x}{|w|=n}
P_\Lambda(B_w) \nn\\
&= \lim_{\Lambda \in \bsLambda} \sumtwo{c \ni 0}{|c|>n} \prod_{j=1}^{|c|}
\e{-\xi(x_{j-1},x_j)} \frac{Z(\Lambda\setminus c)}{Z(\Lambda)} \nn\\
&\leq \sumtwo{c \ni 0}{|c|>n}
\prod_{j=1}^{|c|}
\e{-\xi(x_{j-1},x_j)}. \nn
\end{align}
The first inequality is Fatou's lemma. The last term goes to 0 as $n\to\infty$ since the sum over all cycles containing the origin
converges.
\end{proof}

The typical picture at high temperature is that of \fig\ref{figcyc} $(a)$. Most cycles
involve a unique site and have length 1. When the temperature decreases, cycles lengths
should increase, as depicted in \fig\ref{figcyc} $(b)$.
\bfig
\centerline{$\begin{matrix} \quad \epsfxsize=70mm \epsffile{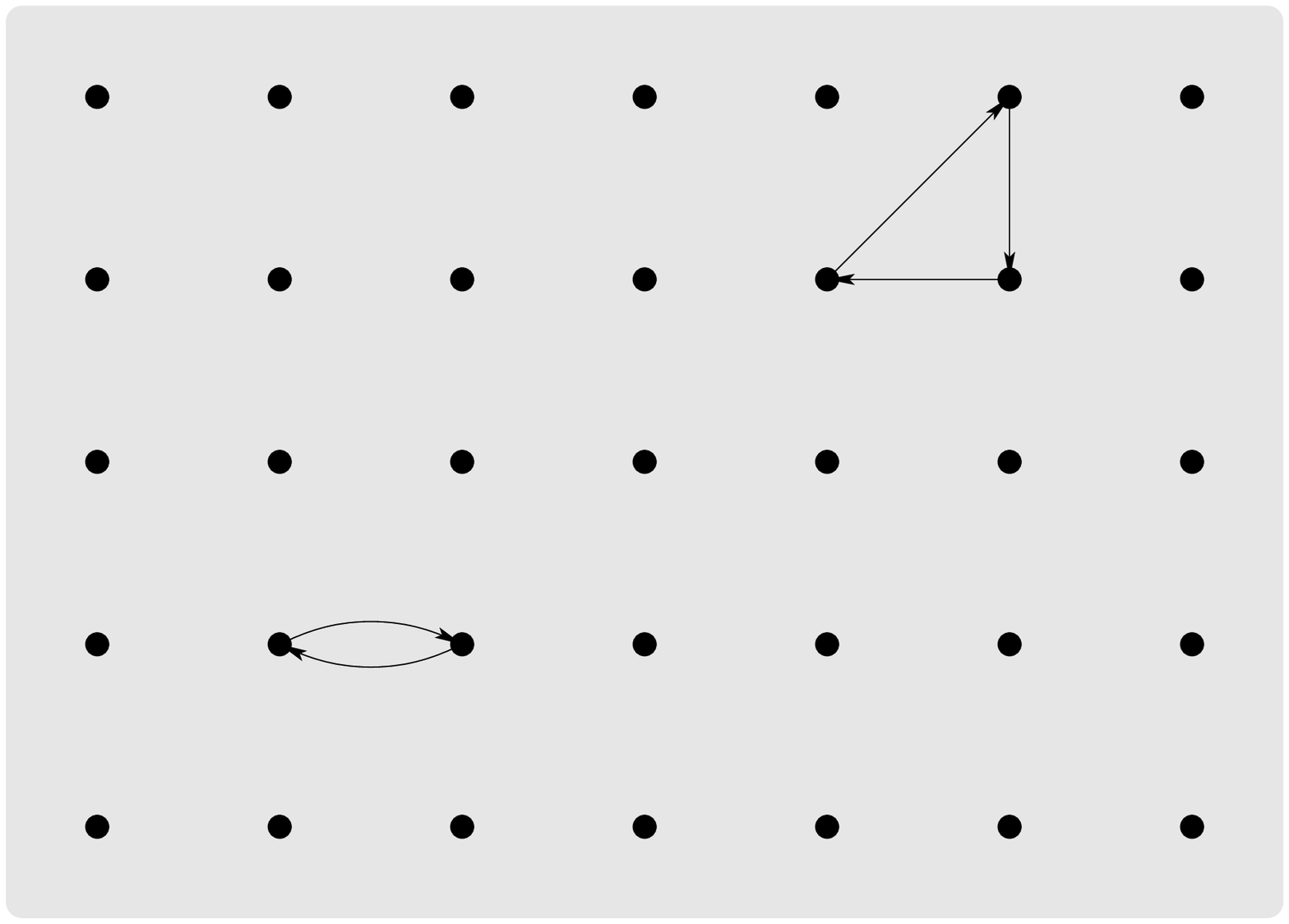} \quad & \quad \epsfxsize=70mm
\epsffile{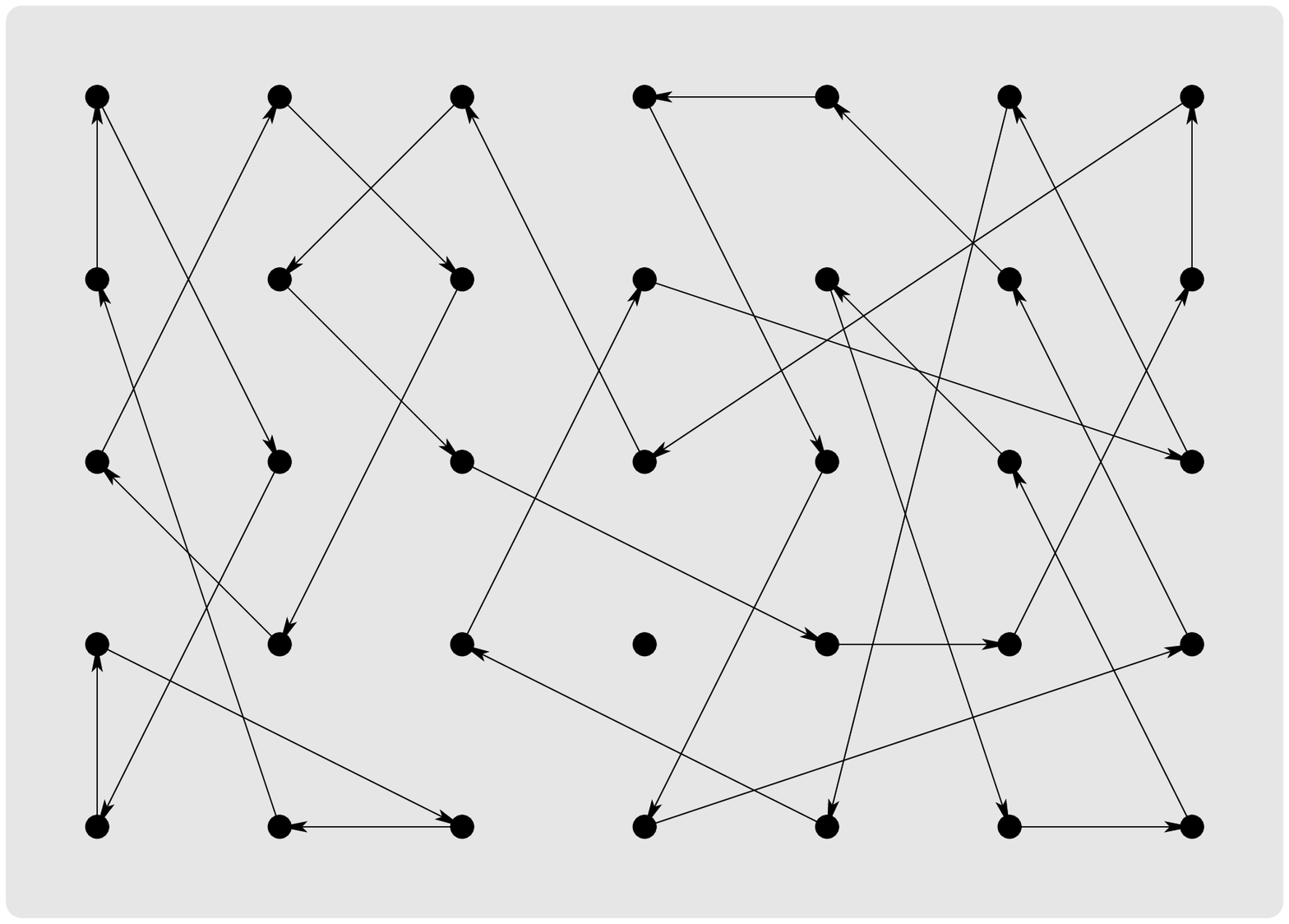} \quad \\ (a) &
(b) \end{matrix}$}
\caption{Expected typical configurations of cycles, $(a)$ at high temperature and $(b)$ at low
temperature.}
\label{figcyc}
\end{figure}
The cycles model resemble that of multiple random walks interacting via exclusions. Assume
for a moment that $\xi$ is given by \eqref{xinn} with $\beta=\infty$, that is, cycles have
nearest-neighbor jumps. One can generate a configuration of cycles by starting at the
origin and doing two self-avoiding random walks in different directions. When they
eventually met, we close this cycle and start another pair of walks from a free site, that
have to avoid the first one. One repeats the procedure until all the sites have been considered.
This actually does not give the same probability distribution on the configurations of
cycles, but one can expect similar behavior. There is a natural question in this process: Is there a chance that after $n$
steps the two legs have not crossed? If the non-crossing probability remains finite when
$n$ goes to infinity, there are infinite cycles. It is actually known that the random walk
is recurrent in dimension 2 and transcient in dimension 3 and higher. Considerably
extrapolating this argument, one obtains an illustration on the fact that BEC occurs only
in dimensions greater or equal to 3. This also suggests the natural conjecture that infinite
cycles do occur in this model at low temperature and $d\geq3$.

\subsection*{Acknowlegments}
I am grateful to R. Moessner and Y. Velenik for a critical reading of the manuscript.

\end{document}